\documentclass[preprint,12pt]{elsarticle}

\usepackage{graphicx}
\usepackage[a4paper, total={6in, 8in}]{geometry}
\usepackage{booktabs}
\usepackage{pdflscape}
\usepackage{longtable}
\usepackage{array}
\usepackage{caption}
\usepackage{amsmath}
\usepackage{titlesec}
\usepackage{algorithm}      
\usepackage{algpseudocode}  
\usepackage[hidelinks]{hyperref}

\titleclass{\subsubsubsection}{straight}[\subsubsection]
\newcounter{subsubsubsection}[subsubsection]
\renewcommand\thesubsubsubsection{\thesubsubsection.\arabic{subsubsubsection}}

\titleformat{\subsubsubsection}[block]
{\normalfont\normalsize\bfseries}
{\thesubsubsubsection}{1em}{}

\titlespacing*{\subsubsubsection}
{0pt}{3.25ex plus 1ex minus .2ex}{1em}

\makeatletter
\def\toclevel@subsubsubsection{4}
\def\l@subsubsubsection{\@dottedtocline{4}{7em}{4em}}
\makeatother

\usepackage{amssymb}
\usepackage{amsmath}

\journal{Journal of Information Security and Applications}

\begin{document}

\begin{frontmatter}



\title{Entity-Aware and Secure Query Optimization in Database Using Named Entity Recognition}


\author[label1]{Azrin Sultana \corref{cor1}} 
	\author[label1]{Hasibur Rashid Chayon}
	\cortext[cor1]{Azrin Sultana. Email address: azrin.sultana.cse@gmail.com}
	
\affiliation[label1]{organization={Department of Computer Science, American International University-Bangladesh},
	addressline={House 83/B, Road 4, Block B}, 
	city={Dhaka},
	postcode={1229}, 
	state={},
	country={Bangladesh}}

\begin{abstract}
Cloud storage has become the backbone of modern data infrastructure, yet privacy and efficient data retrieval remain significant challenges. Traditional privacy-preserving approaches primarily focus on enhancing database security but fail to address the automatic identification of sensitive information before encryption. This can dramatically reduce query processing time and mitigate errors during manual identification of sensitive information, thereby reducing potential privacy risks. To address this limitation, this research proposes an intelligent privacy-preserving query optimization framework that integrates Named Entity Recognition (NER) to detect sensitive information in queries, utilizing secure data encryption and query optimization techniques for both sensitive and non-sensitive data in parallel, thereby enabling efficient database optimization. Combined deep learning algorithms and transformer-based models to detect and classify sensitive entities with high precision, and the Advanced Encryption Standard (AES) algorithm to encrypt, with blind indexing to secure search functionality of the sensitive data, whereas non-sensitive data was divided into groups using the K-means algorithm, along with a rank search for optimization. Among all NER models, the Deep Belief Network combined with Long Short-Term Memory (DBN-LSTM) delivers the best performance, with an accuracy of 93\% and precision (94\%), recall, and F1 score of 93\%, and 93\%, respectively. Besides, encrypted search achieved considerably faster results with the help of blind indexing, and non-sensitive data fetching also outperformed traditional clustering-based searches. By integrating sensitive data detection, encryption, and query optimization, this work advances the state of privacy-preserving computation in modern cloud infrastructures.

\end{abstract}


\end{frontmatter}

\section{Introduction}
Cloud computing enables online data storage, analysis and access services to various clients, significantly lowering the cost of computing \cite{r1}. In the cloud environment, data outsourcing methods have become immensely popular \cite{r2}. As a result, an increasing number of individuals and companies are eager to outsource their data to the cloud due to the rapid development of NoSQL databases, which can store enormous amounts of unstructured and semi-structured data \cite{r3} \cite{r4}. However, data outsourcing conveys accessibility to data users, yet it also inevitably entails the risk of data breach. Untrustworthy service providers or attacker might steal user's personal information, such as email, address, and phone numbers, credit card number and sell to third parties. More importantly, attackers who target a cloud server can retrieve customers' sensitive personal information \cite{r5}\cite{r6}\cite{r7}. Customers' personally identifiable information, such as home addresses, phone numbers, emails, and Social Security numbers, is involved in approximately half (46\%) of all breaches \cite{r8}. According to an IBM estimate from 2023, cloud-stored data is involved in 82\% of data breaches. The cost is higher than average, at \$4.75 million, and 39\% of breaches involve several environments \cite{r9}. 

Data encryption substantially alleviates the risk of privacy breaches by preventing cloud providers from learning the data they keep. Nevertheless, it is not scalable to manually identify and isolate such personal data or to reuse rule-based procedures, particularly when the data is unlabeled or unstructured \cite{r10}\cite{r11}. This makes it essential to rely on intelligent automation capable of handling diverse and unlabeled data, which is often encountered in this domain. Integrating NER into this system facilitates efficient identification and protection of sensitive attributes automatically, enabling secure data processing without compromising performance \cite{r12}\cite{r13}. NER is a vital Natural Language Processing (NLP) technology that allows machines to identify and categorise the essential parts of text. NER enables the machine to categorise text into specified categories, which helps the system automatically locate sensitive content in NoSQL queries \cite{r14}\cite{r15}. Automatic extraction of sensitive entities is the first step toward implementing privacy-driven approaches, such as encryption, indexing, and access control. Ensuring large-scale privacy without NER would be an impossible task, leading either to over-encryption of non-sensitive data or missing protection for sensitive data, which would negatively impact both security and query performance.

During a data breach or unauthorised access, encrypted information such as bank records, personally identifiable information, or medical records is inaccessible to intruders, as they do not have access to the key \cite{r16}. There is, however, a dreadful limitation to this added security feature: encrypted data cannot be accessed via common query methods \cite{r17}. As NoSQL databases are schema-less, normal queries such as searching, sorting, and range queries on encrypted values are challenging without handling the ciphertext through processing \cite{r18}. As a result, the efficiency of the databases is impeded due to the encryption.  So, the growing need is to do query optimisation that operates directly on encrypted data \cite{r19}. Furthermore, without optimisation, querying a database, especially systems that require real-time analysis, fast data transmission, and a privacy-aware system, can have serious consequences \cite{r20}. In these contexts, balancing encryption with intelligent indexing, clustering, and ranking mechanisms ensures that security does not compromise system usability. Therefore, developing privacy-preserving query optimisation methods is a critical research direction for modern NoSQL systems, especially as organisations seek to protect sensitive data while maintaining fast and accurate query performance at scale \cite{r21}\cite{r22}.

Whereas most projects optimise query performance, enhance database security, or identify private data alone, the novelty of our proposed model lies in its ability to do both, preserving user data privacy by automatically identifying and optimising. Based on our literature review, no existing work applies NER to identify sensitive information, encrypt it before sending to the database, and utilise clustering to optimise search for non-sensitive information. Working with databases, such as encrypting information and optimising, requires in-depth knowledge about databases and a system-level understanding of how secure data flows can coexist with intelligent information retrieval \cite{r23}. Because these fields rarely overlap in standard research or industry solutions, and need a vast amount of knowledge, no frameworks, tools, or best practices exist to support such an effort. The lack of availability of a sensitive data-labeled real-world dataset \cite{r24} makes it difficult to work in this domain. Therefore, it makes it difficult to train, test, and benchmark systems that aim to both identify sensitive data and encrypt it securely, while optimising queries. These gaps in existing research highlight the urgent need for a framework that not only secures sensitive data by automatically detecting it but also enables intelligent query optimisation over both encrypted and plain data.

In this study, to address this need, we propose a comprehensive privacy-aware framework to optimise queries on the NoSQL database, utilising an NER model, encryption, and a secure indexing technique. The primary objective is to automatically identify sensitive information, including email addresses, phone numbers, and credit card numbers, utilising a trained NER model. When identified, these sensitive entities are encrypted using advanced encryption algorithms. To support efficient queries on encrypted data, we use blind indexing. To further enhance retrieval accuracy and user experience, we apply Term Frequency-Inverse Document Frequency (TF-IDF)-based vectorising and K-means clustering for ranked results for non-sensitive information. The purpose of our proposed study is to strike a balance between strong privacy guarantees and high query efficiency, which are suitable for privacy-sensitive domains such as finance, healthcare, and corporate communication. Through experimental validation on the Enron dataset, we demonstrate the effectiveness of our proposed approach in terms of security, speed, and retrieval relevance.

OUR CONTRIBUTION
\begin{itemize}
	\item A designated NER-driven automated pipeline for detecting and classifying sensitive data (e-mail, phone numbers, credit card numbers) within the unstructured NOSQL data.
	\item In this model, the blind index, which enables effective and secure queries to encrypt data, was used for efficient retrieval of sensitive data without decrypting.
	\item Integrated AES-Symmetrical encryption that maintains data confidentiality and security, to encrypt text to protect sensitive data safely.
	\item In this model, the blind index, which enables effective and secure queries to encrypt data, was used for efficient retrieval of sensitive data without decrypting
	\item For optimising non-sensitive information search, a TF-IDF-based vectorising system and K-Means clustering were included to increase the relevance and purpose of the query optimisation.
\end{itemize}

The remaining structure of this paper is as follows. The literature review is given in Section II, and the methodology of our proposed model is described in Section III, where we elaborate on our proposed system in detail and discuss the optimization technique. Experimental performance analysis is presented in Section VI, and the discussion is in Section V. Finally, we give the conclusion and future direction of this paper in the conclusion section.

\section{LITERATURE REVIEW}

The massive increase in data outsourcing and cloud storage has revolutionized the demand for secure and efficient data retrieval as a top field in today's data management. In response to this, prior research proposed various encryption schemes and indexing techniques to enable security and fast search over encrypted data. The existing approaches, however, typically rely on manual identification of sensitive information or on encrypting the entire database, which introduces human error and impedes scalability. According to our model, sensitive information detection via NER not only reduces errors but also allows for more effective encryption and indexing. Previous works propose independent algorithms to maximize encryption approaches; however, they have some limitations. None makes any attempt at integrating sensitive data identification, encryption, and database optimization into a comprehensive framework—something our model addresses to enhance system performance significantly.

\subsection{IDENTIFYING NAMED ENTITY RECOGNITION}

Dias et al. \cite{r25} proposed a model capable of extracting and classifying Portuguese-language-sensitive data from the corpus. For the study, the Portuguese dataset annotated with entity classes, the HAREM dataset, and the SIGARRA News Corpus were used. For testing the model, the DataSense NER Corpus was used. The recognition of named entities is based on three submodules: a rule-based approach for identifying sensitive information; lexicon-based analysis to identify socio-economic information; and ML models for more ambiguous entities where rule-based and lexicon-based methods are insufficient. Conditional Random Field, Random Forest Bidirectional, and LSTM were used to train and test the dataset, with 5-fold cross-validation and 80 epochs to alleviate overfitting; the dropout rate was 0.68. The model had an F1 score of 83.01\%. 

By using numerous DL models, Sun et al. \cite{r26} proposed a prediction model for NER recognition for natural hazards and analyzed the models based on three steps, which include pretraining, extracting features, and decoding. For this model, the Wanfang Database’s research articles, comprising 12,387 papers with only the title and abstract, were compiled. For annotating the dataset, four authors worked together. In this paper, Bidirectional Encoder Representations from Transformers (BERT), ALBERT, and XLNet were trained on large-scale, unlabeled natural hazard corpora. Beginning, Inside, and Outside (BOI) labels were used to label the entities, whereas ALBERT was used for sentence-order prediction. The proposed model accurately identifies contextual words using XLNET. BiLSTM and BiGRU were used for extracting significant features. Combining XLNET, BiLSTM, and Conditional Random Field (CRF), the model achieved the highest performance, with a precision of 92.80\%, recall of 91.74\%, and an F1-score of 92.27\%. 

To identify sensitive information in unstructured documents, Ahmed et al. \cite{r27} applied DL, machine learning, and rule-based methods. For the study, two public datasets were used: one was a text-sensitive, context-dependent dataset collected from Kaggle with 1 million instances, and 400 instances were selected due to a lack of labelling. On the other hand, the image dataset was collected from three sources, comprising 300 images labelled as sensitive or non-sensitive. By using leave-one-out cross-validation, they validated the model. Three different types of model, machine learning-based, rule-based, and DL-based, were used, and their results were compared. To verify missing letters or digits in images, the Levenshtein distance metric was used to measure the distance between words and keywords. Among these three models, the DL-based model outperformed the others, achieving an accuracy of 95\%. 

To extract patient-reported data from unstructured text, \cite{r28} used a weakly supervised model trained with a medical NER model across eight languages. The first step was merging several databases of medical texts to create a complete corpus. To mitigate the class imbalance problem, data augmentation was applied to the dataset, which also helped alleviate overfitting. The fine-tune model achieved its best performance in English, with an F1 score of 80.07\%, followed by German, Spanish, and Portuguese, all of which scored over 77\%. In underrepresented languages, Slovak, Italian, and Polish have promising results with 75\%.

Five components—feature extraction, visual, textual, and multimodal feature fusion- are employed in the paper \cite{r29} to propose a high-quality protocol for constructing privacy data in Android privacy-sensitive datasets for power equipment. The data were web-scraped and annotated using the BIO tagging scheme. The model consists of three main components: text feature extraction, character embedding construction, and the CRF Viterbi decoder. BiLSTM-based feature extraction with multi-head attention achieved better performance than baseline methods.

Using adversarial training and external context retrieval, \cite{r30} proposed an NER model for identifying automotive defect messages in Chinese. The model uses a search engine to retrieve the character–word fusion embedding representation and an enhanced keyword-extraction approach to generate externally relevant texts semantically connected to the original input phrase.  After that, the attention mechanism is employed to fuse the character–word fusion embedding representations. Lastly, a BiLSTM is used to extract long-distance semantic information, and a CRF is used to acquire the recognition result. Experiments on both domain-specific and general NER datasets (e.g., Weibo, Resume) demonstrate that the proposed model outperforms existing models, achieving F1-scores of 89.41\% on automotive datasets, which indicates strong generalization and robustness.

More articles related to NER detection summary are provided in Table \ref{tab:ner_studies}.

\begin{longtable}{|p{1.5cm}|p{4cm}|p{4cm}|p{4cm}|}
	\caption{Summary of Named Entity Recognition Studies}
	\label{tab:ner_studies} 
	\\ \hline
	\textbf{Publish Year} & \textbf{Objective} & \textbf{Dataset} & \textbf{Algorithms} \\ \hline
	\endfirsthead
	
	\hline
	\textbf{Publish Year} & \textbf{Objective} & \textbf{Dataset} & \textbf{Algorithms} \\ \hline
	\endhead
	
	Wilkho et al. 2025 \cite{r37} & Develop a Chinese NER method for news texts & News-domain Chinese NER corpus & Transfer learning (ERNIE) and Soft-Lexicon embeddings \\ \hline
	
	Gao et al. 2025 \cite{r38} & Develop a tourism-specific NER model that predicts entity boundaries to effectively recognize nested and long entities in the tourism domain & Three tourism datasets combining public and private sources & BiLSTM \\ \hline
	
	H. Cho et al. 2019 \cite{r39} & Extract information from biomedical articles & NCBI Disease Corpus, BioCreative II Gene Mention, Chemicals-Disease Relationship corpus & BiLSTM, CRF, BERT, CLSTM \\ \hline
	
	Zhou et al. 2025 \cite{r40} & DL-based NER for extracting information from construction incident reports & Construction site incident reports & BERT, BERT-LSTM-CRF, BERT-LSTM \\ \hline
	
	Serere et al. 2025 \cite{r41} & Dynamic NER model detecting authors' stance and location references & Twitter corpus of 4,098 posts (Harvard Dataverse) & LLM \\ \hline
	
	G. et al. 2025 \cite{r42} & Dynamic NER for distinguishing current vs. remote locations in cybersecurity tweets & Custom Twitter dataset annotated using grammatical rules & Fine-tuned LLM \\ \hline
	
	Wu et al. 2018 \cite{r43} & Extract conceptual clinical text & MIMIC II corpus & RNN, CNN \\ \hline
	
	Zhang et al. 2025 \cite{r44} & Security- and privacy-informed NER for entity extraction & Data from Whois, Censys, and Shodan & BERT, GLM4-9B, Mixtral-8B, LLaMa3-8B models \\ \hline
	
	Wang et al. 2024 \cite{r45} & Few-shot multi-view NER for cybersecurity threat intelligence & Threat Intelligence Dataset & BERT, BiLSTM+CRF \\ \hline
	
	Demirol et al. 2025 \cite{r46} & Extract cyber threat entities (actors, malware, campaigns, targets) & 100 PDF reports + 540 HTML pages, manually annotated & BERT \\ \hline
	
\end{longtable}

\subsection{SECURING AND OPTIMIZING THE DATABASE}
For optimized database operations, encrypted data must be searchable and sortable, which means some information must be exposed. For this purpose, Wiese et al. \cite{r31} developed a framework that achieves maximum security, functionality, and performance in property-preserving encryption for unmodified wide column stores in NoSQL databases. They employed index-based encryption schemes to scale the database. To execute schemas on encrypted data, a Java Cryptography Extension KeyStore was used to store a random number of keys retrieved using a custom label. Searchable encryption is only performed on strings, while order-preserving encryption is performed on strings or numerical values, meaning that indexes are preserved per column. The proposed model measured the communication time between the encrypted query and the retrieval of the same results.

Incorporating a client’s authorization data into search tokens and indexes, Du et al. \cite{r32} proposed a multi-client symmetric encryption schema. The proposed model enabled multiple clients to execute Boolean queries over an encrypted database, restricting search capabilities to specific keywords, thereby providing double-layer security. When a malicious user queries with keywords outside their authorization and attempts to access unauthorized content, the system blocks that request. The model was evaluated in terms of computation and communication costs. Supports complex search queries involving Boolean operations (e.g., AND, OR, NOT), enabling more expressive search conditions.

He et al. \cite{r33} proposed a fast, searchable encryption scheme with ranked search, and a clustering algorithm was adopted to cluster the documents. For clustering, a keyword conversion method was used to convert documents and queries into vectors. A tree-building algorithm was applied to produce an index tree for each document set, with relevance scores between the query and the index tree roots, and the highest correlation score was selected using symmetric encryption. The comparative results suggested that query efficiency was sublinear in document size.

A secure, ranked search of encrypted cloud data is proposed by Elizabeth et al. \cite{r34}. n encrypted searchable index for keywords and a trapdoor corresponding to the search keyword are constructed. Using the encrypted score index, it validated and calculated scores for the query keyword and ranked relevant queries from the document, along with verifiable parameters. For trapdoor and index generation, the Paillier cryptosystem was used, ensuring that neither the third party nor the provider has any knowledge of the ranking. The model dynamically updated those encrypted fields without rebuilding the entire index.  To verify query results, the scheme provided a verifiable search using keyed hashes. Simulations on the Enron dataset indicate that the proposed scheme is both semantically secure and preserves correctness and privacy. The authors make significant contributions to the field of verifying generated output and updating query scenarios. However, the more complex the system, the more space it requires; how they handle space complexity is not discussed. Moreover, information that is regarded as sensitive is not mentioned, and is designed only for a single data owner, and how this system works.

A multi-keyword searchable encryption scheme is presented \cite{r35} for secure and efficient search, especially for supporting multi-keyword queries without compromising security. Introduced a five-phase scheme: key generation, index building, trapdoor generation, query processing, and result retrieval. The core innovation lies in the use of a probability-based trapdoor and keyword vector to enhance search security and query flexibility. The resistance to the probability trapdoor construction safeguards the scheme’s security.  The authors prove that the probabilistic design prevents reverse-engineering of search patterns and keyword exposure. 

\cite{r36} discusses the growing demand for secure and privacy-preserving searchable encryption in the cloud for multiple users and keywords by using a two-server model and a private subset decision mechanism. It ensures data and query privacy, constant-size ciphertexts and trapdoors, and multi-user access in cloud environments. The scheme utilises cryptographic primitives for safe bit decomposition and a trapdoor key cryptosystem to support secure search operations. Simulation-based security is also analyzed, ensuring strong privacy against non-colluding, semi-honest entities. Experimental results suggest that the proposed system's trapdoor generation, ciphertext size, and query efficiency surpass those of previous approaches, especially at scale.

Many studies have been tackling sensitive information detection using methods ranging from rule-based systems to DL approaches. While they worked well with the assistance of NER and transformer models, they were not generalizable and were computationally intensive. On the other hand, encryption-focused models, which enhanced indexing for searching, failed to adequately address privacy scalability and granularity. Meanwhile, other authors have proposed secure, ranked, and verifiable search models; however, these were limited by complexity, a single-user architecture, and inconsistent definitions of sensitive data. Nonetheless, no research has been conducted on the automatic detection of sensitive information and NoSQL database optimization in tandem. To overcome the previously mentioned drawbacks and bridge the gap between database optimization and the automatic identification of sensitive information using an NER model, we propose an efficient NER model that is computationally inexpensive, as we have not utilized any advanced models. However, our model has not sacrificed performance. We have only encrypted sensitive information, rather than the entire database, which is the most efficient and fastest approach. To secure the database, we have optimized it by implementing two-layer security and utilizing indexing and clustering of non-sensitive data for efficient retrieval.

\section{METHODOLOGY}
This chapter describes the process adopted to design and experiment with the proposed privacy-conscious data identification and optimization of the database framework. The approach integrates an NER model to automatically identify sensitive information, followed by encryption and indexing for secure query processing. Additionally, clustering, vectorized form search, and ranked search are performed on non-sensitive data to enhance retrieval efficiency. There is an elaborate explanation of each stage of the system, including the preprocessing step, encryption step, and performance metrics used for assessment. For this research, the proposed system combined NER with database security and query optimization, which is a benchmark in this field. Using NER for automatically identifying sensitive information, such as contact numbers, email addresses, social security numbers, and credit card numbers, eliminates the extra work required for manually handling sensitive information. In addition to our proposed approach, it also efficiently optimized queries without compromising security.
\subsection{DATASET DESCRIPTION}
The dataset for this is collected from Mendeley \cite{r47}, which has anonymized personally identifiable information. The dataset imitates various personal entities naturally found in financial contexts. It is intended to support the development and evaluation of sensitive data. Each entry feigns real-world information. The dataset has 45,000 instances with 10 attributes, namely names, social security numbers, credit card numbers, phone numbers, email addresses, physical addresses, company names, text, and URLs. For this research, we have only taken the text attribute. In most text attributes of the dataset, sensitive information is present. We have carefully collected and meticulously annotated this data to enable precise identification. These identified entities are subsequently utilized to detect sensitive information within NoSQL databases. Table \ref{tab:sensitive_entities} provides the details of the sensitive information collected from this dataset. The text is composed of an average length of 1,357 characters, ensuring a comprehensive and detailed exposition that facilitates clear understanding and precise information. For verifying our database optimization model, we have used enron dataset, which have sensitive information such as email addresses, and non sensitive information as well. Which is ideal to verify the query optimization system.

\begin{table}[htbp]
	\centering
	\caption{Entity details from the dataset}
	\label{tab:sensitive_entities}
	\begin{tabular}{p{3cm}p{3cm}p{4cm}}
		\toprule
		\textbf{Category} & \textbf{Entity Name} & \textbf{Sensitive Data Included} \\
		\midrule
		Personal Identification Number & PASSPORT & Passport Number \\
		& PHONE & Telephone Number \\
		& CREDIT\_CARD & Credit card number \\
		& EMAIL & Email address \\
		\midrule
		Socio-Economic Information & SSN & Social Security Number \\
		\midrule
		Others & URL & Link \\
		& PAYMENT & Money \\
		\bottomrule
	\end{tabular}
\end{table}

\subsection{MODEL DESCRIPTION}

Our proposed research is divided into three segments.  In the first stage, after user send a query, identified sensitive information such as phone number, email address, social security number, link, and credit card number using DL and NLP, by our NER model and then we used encryption techniques for encrypting the identified sensitive data, as database attribute encryption is different from normal encryption, which is the second stage, privacy. We must ensure that encrypted values remain fully searchable and capable of supporting conditional queries, even while their content remains concealed, thereby enabling efficient and secure data utilization. The final stage of our proposed model involves applying database optimization techniques to enhance performance, speed, and security, even though sensitive data is encrypted and then stored in the database. 
\subsubsection{NAMED ENTITY RECOGNITION MODEL}

\begin{figure}[htbp]
	\centering
	\includegraphics[width=\textwidth]{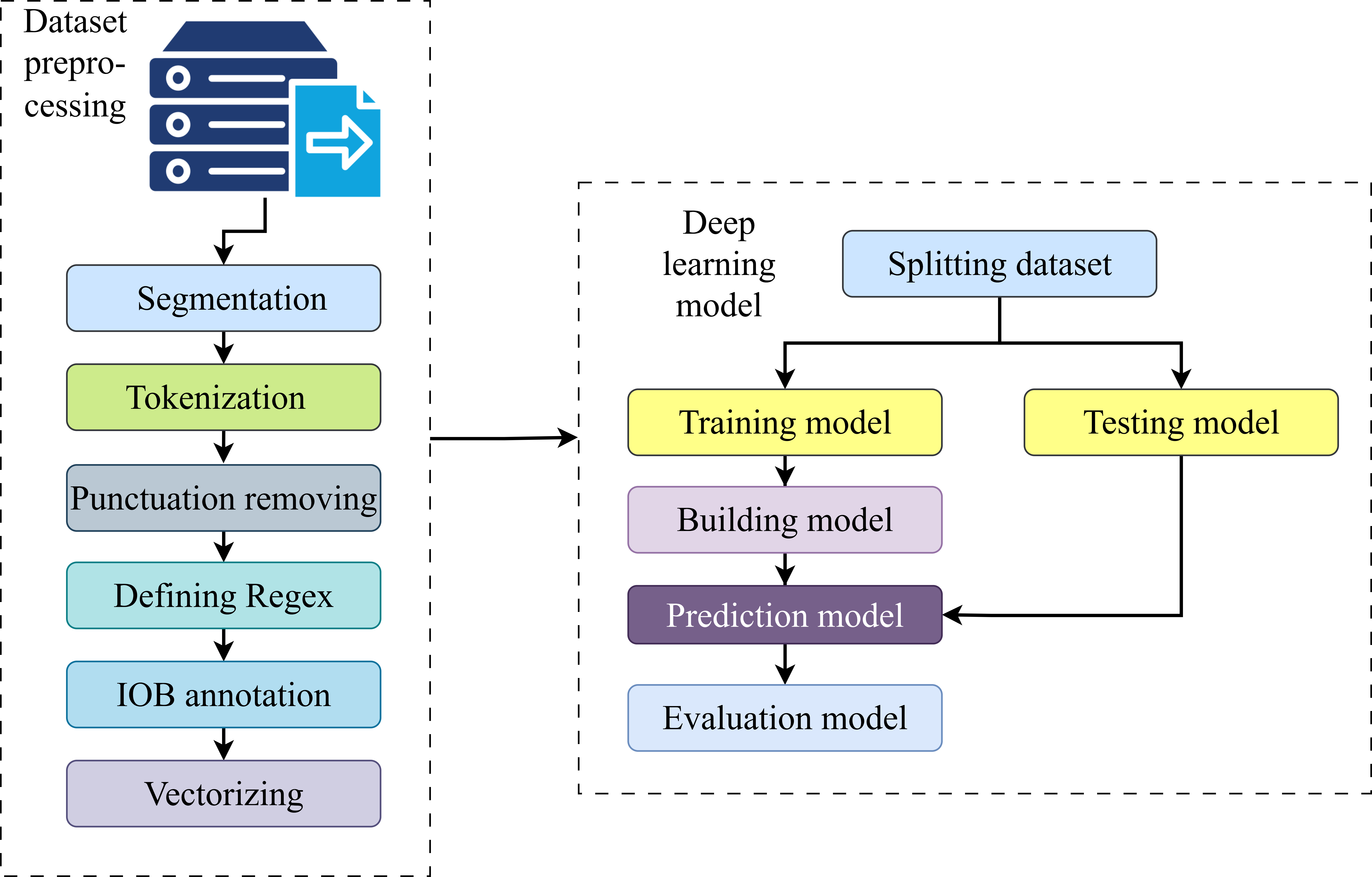}
	\caption{Named entity recognition model flow diagram}\label{nermodel}
\end{figure}

The NER model is used to identify sensitive information from unstructured data. This is the first step of our proposed model. From Figure \ref{nermodel}, the dataset is preprocessed using segmentation, tokenization, punctuation removal, defining a regular expression (regex), IOB format annotation, and vectorization. After preprocessing, the DL model is used to train, test, and validate the model. Each step is discussed in detail.

\subsubsubsection{TOKENIZATION}
Tokenization involves splitting the raw text into smaller chunks of words or sentences, which are referred to as tokens. When a sentence is split into words, it is called word Tokenization, and sentence tokenization refers to splitting a sentence into individual sentences. Primarily, the space between words is determined for accomplishing the word tokenization, whereas characters like exclamation points and periods are used for sentence tokenization \cite{r48}. For our proposed model, word tokenization is employed to break down sentences into individual words, as shown in Table \ref{tab:tokenization}.

\begin{table}[tbp]  
	\centering
	\caption{Word tokenization of the text}
	\label{tab:tokenization}
	\begin{tabular}{p{0.45\linewidth} | p{0.45\linewidth}}
		\toprule
		\textbf{Sentence} & \textbf{Words} \\
		\hline
		\midrule
		To the members of Andrade LLC & 'To', 'the', 'members', 'of', 'Andrade', 'LLC' \\
		\hline
		Following the annual compliance & 'Following', 'the', 'annual', 'compliance' \\
		\hline
		Independent auditors' report on & 'Independent', 'auditors'', 'report', 'on' \\
		\hline
		We have conducted a thorough search. & 'We', 'have', 'conducted', 'a', 'thorough', 'search' \\
		\hline
		To whom it may concern, this & 'To', 'whom', 'it', 'may', 'concern', 'this' \\
		\hline
	\end{tabular}
\end{table}

\subsubsubsection{REMOVING PUNCTUATION MARKS FROM TEXT}
Removing punctuation marks from the text is a crucial process in NLP. Punctuation marks, namely the dot, comma, colon, and question mark, do not have any significant meaning; however, they can sometimes skew the results, and the model can be trained with noise due to word frequency and embeddings \cite{r49}. Therefore, removing punctuation makes the text simpler, reshuffling the analysis by plummeting the complexity and inconsistency within the data \cite{r50}. In this model, punctual marks are removed during the text preprocessing. Removing all punctuation marks affects our data because many of our sensitive words contain punctuation marks, such as periods, hyphens, and commas. In email, there are dots, and a comma can be used in the payment amount. Additionally, hyphens can be used in social security numbers, passports, and credit card details. So, we made an algorithm that only removed punctuation marks at the end of the words, so that our intended values won’t be hurt. Our punctuations discard details are given in Table \ref{tab:text_cleaning}.
\begin{table}[tbp]  
	\centering
	\caption{Cleaned words after removing punctuation}
	\label{tab:text_cleaning}
	\begin{tabular}{p{0.45\linewidth} | p{0.45\linewidth}}
		\toprule
		\textbf{Words} & \textbf{Cleaned words} \\
		\hline
		\midrule
		jonathon@gmail.com, & jonathon@gmail.com \\
		\hline
		10,500. & 10,500 \\
		0987345567689, & 0987345567689 \\
		\hline
		(017123455688) & 017123455688 \\
		\hline
	\end{tabular}
\end{table}
\subsubsubsection{DEFINING REGULAR EXPRESSIONS}	
A Regular expression (Regex) is an arrangement of words that is used for searching patterns \cite{r51}. We applied a Regular Expression to identify sensitive information from the texts. First, we scrupulously observed the pattern of sensitive information present in the text, and based on these patterns, we developed a regular expression (regex) for identifying them. We applied these regexes after removing punctuation marks from the text. Table \ref{tab:regex} describes the regular expression used for determining values.

\begin{table}[tbp]
	\centering
	\caption{Regular expressions for identifying dataset sensitive information}
	\label{tab:regex}
	\begin{tabular}{p{0.23\linewidth} | p{0.17\linewidth} |p{0.55\linewidth}}
		\toprule
		\textbf{Name} & \textbf{Entity} & \textbf{Regular Expression} \\
	
		\midrule
		Email address & EMAIL & \verb+[a-zA-Z0-9.%+-]+@[a-zA-Z0-9.-]+\.[a-zA-Z]{2,}+ \\

		Phone number & PHONE & \texttt{\textbackslash+(\textbackslash d+)(?:\textbackslash s?(\textbackslash d\{6,\}))?} \\
	
		Social security number & SSN & \verb+\b\d{3}-\d{2}-\d{4}\b+ \\
	
		Payment amount & MONEY & \verb+\$\d{1,3}(?:,\d{3})*(?:\.\d{2})?+ \\
	
		Credit card number & CREDIT\_CARD & \verb+\b\d{13,16}\b+ \\
	
		Link & URL & \verb+https?://[^\s]++ \\
		\bottomrule
	\end{tabular}
\end{table}

\begin{figure}[htbp]
	\centering
	\includegraphics[width=\textwidth]{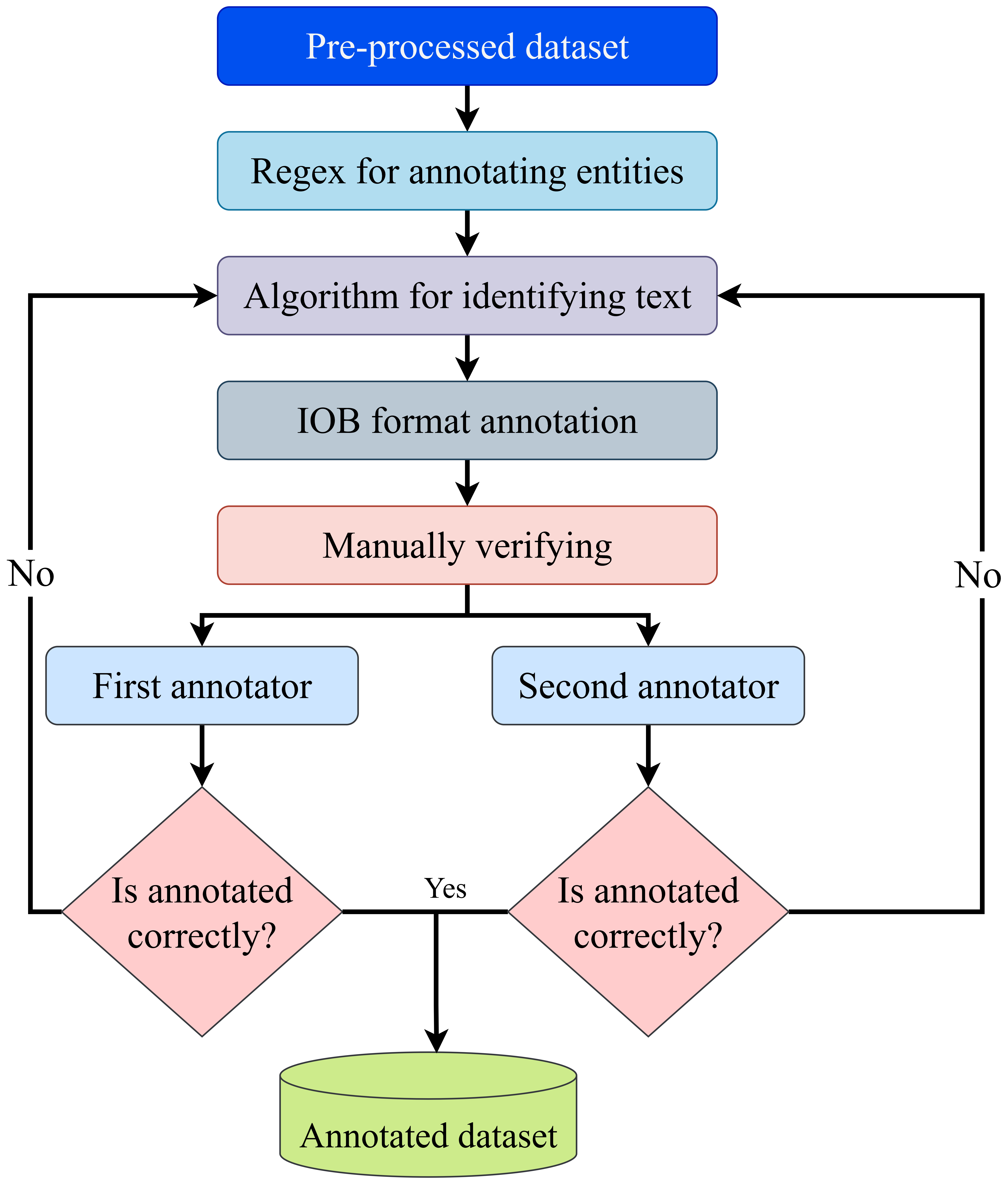}
	\caption{ IOB format annotation flow diagram}\label{figure:flowiob}
\end{figure}

\subsubsubsection{IOB FORMAT DATASET ANNOTATION} 
\begin{figure}[htbp]
	\centering
	\includegraphics[width=\textwidth]{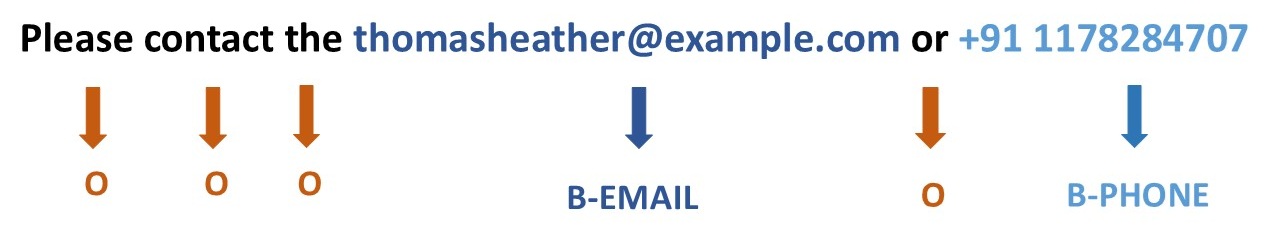}
	\caption{ IOB format annotation}\label{figure:IOB_annotation}
\end{figure}
Dataset annotation is the method of labeling the data into predefined classes. It is a crucial process, particularly for supervised learning, and can be done for diverse types of datasets. Data annotation plays a crucial role in evaluating the performance of models, as it provides the target labels. At least two individuals should supplement the annotation process to validate the trustworthiness of the annotations \cite{r52} \cite{r53}. There are several techniques for annotating the corpus. IOB refers to inside, outside, and beginning of the tag, which are all part of it. Figure \ref{figure:IOB_annotation} depicts the IOB format. As algorithms such as LSTM, CNN, and transformer models become more precise in understanding IOB for sequence labeling, they can train models with greater accuracy. Additionally, it is easier to verify the proposed model by using performance metrics when the dataset entities are in IOB format \cite{r54}. For this reason, we employed this technique to annotate our dataset. After annotating based on regular expressions, we manually verified the accuracy of the annotations. If we find any mishandling or errors in annotating text, we resolve the issue, modify the algorithm's logic for annotating texts, and then re-run Algorithm 1 for annotation. The flow diagram of dataset annotation is shown in Figure \ref{figure:flowiob}, which depicts the entire annotation process using the IOB format. Datasets are preprocessed, and then an algorithm is applied to identify sensitive texts, which are formatted using the IOB scheme. Finally, we manually checked for any incorrect identification and then rewrote the algorithm \ref{alg:ner_labeling}.

\begin{algorithm}
	\caption{Token Labeling for Named Entity Recognition}
	\label{alg:ner_labeling}
	\begin{algorithmic}[1]
		\Procedure{label\_tokens}{tokens, patterns}
		\State labels $\gets$ list of ``O'' with length |tokens|
		\State active\_entity $\gets$ None
		
		\For{$i = 0$ \textbf{to} |tokens|-1}
		\State token $\gets$ tokens[i]
		\State entity\_found $\gets$ false
		\For{entity, pattern \textbf{in} patterns}
		\If{token matches pattern}
		\If{active\_entity $\neq$ entity}
		\State labels[i] $\gets$ ``B-'' + entity
		\State active\_entity $\gets$ entity
		\Else
		\State labels[i] $\gets$ ``I-'' + entity
		\EndIf
		\State entity\_found $\gets$ true
		\State \textbf{break}
		\EndIf
		\EndFor
		\If{not entity\_found}
		\State active\_entity $\gets$ None
		\EndIf
		\EndFor
		
		\State \Return labels
		\EndProcedure
	\end{algorithmic}
\end{algorithm}

\subsubsubsection{WORD TO VECTOR CONVERSION }	
Machine learning algorithms do not understand the texts. Therefore, we need to convert text into numeric forms. Word embedding is a method used to represent words and documents in corresponding numeric values, known as word vectors. A word vector characterizes a word in a lower-dimensional space, allowing words with comparable meanings to have a similar representation. By using this method, features are extracted from the text and then input into an ML model, preserving both syntactical and semantic information \cite{r55} \cite{r56}. Word2Vec is a neural method for producing word embeddings. It is a widespread technique in NLP to represent words as continuous vector spaces, which collect the semantic relationships between words by mapping them to high-dimensional vectors, and every word is allocated a vector \cite{r57}. For our study, we utilized Word2Vec to transform words into vectors. Compared to other traditional methods, it produces dense groups of semantic words that are closer together, which helps reduce complexity and decrease training time. Less complexity, reduced training, and effective representation of dense vectors are the most suitable features for our dataset classification for identifying sensitive information
\subsubsubsection{TRAINING NER MODEL}	 
After the preprocessing is done and we have a machine-understandable form of our texts, we feed them into deep learning models by splitting them into an 80:20 ratio. We trained the model using DL algorithms and also hybrid models. We used three DL algorithms and built hybrid models, which are discussed below in detail in the algorithm section. After training is completed, we have tested our model, and the performance details are given in the results section
\subsubsection{SECURING AND OPTIMIZING THE DATABASE} 
\begin{figure}[htbp]
	\centering
	\includegraphics[width=\textwidth]{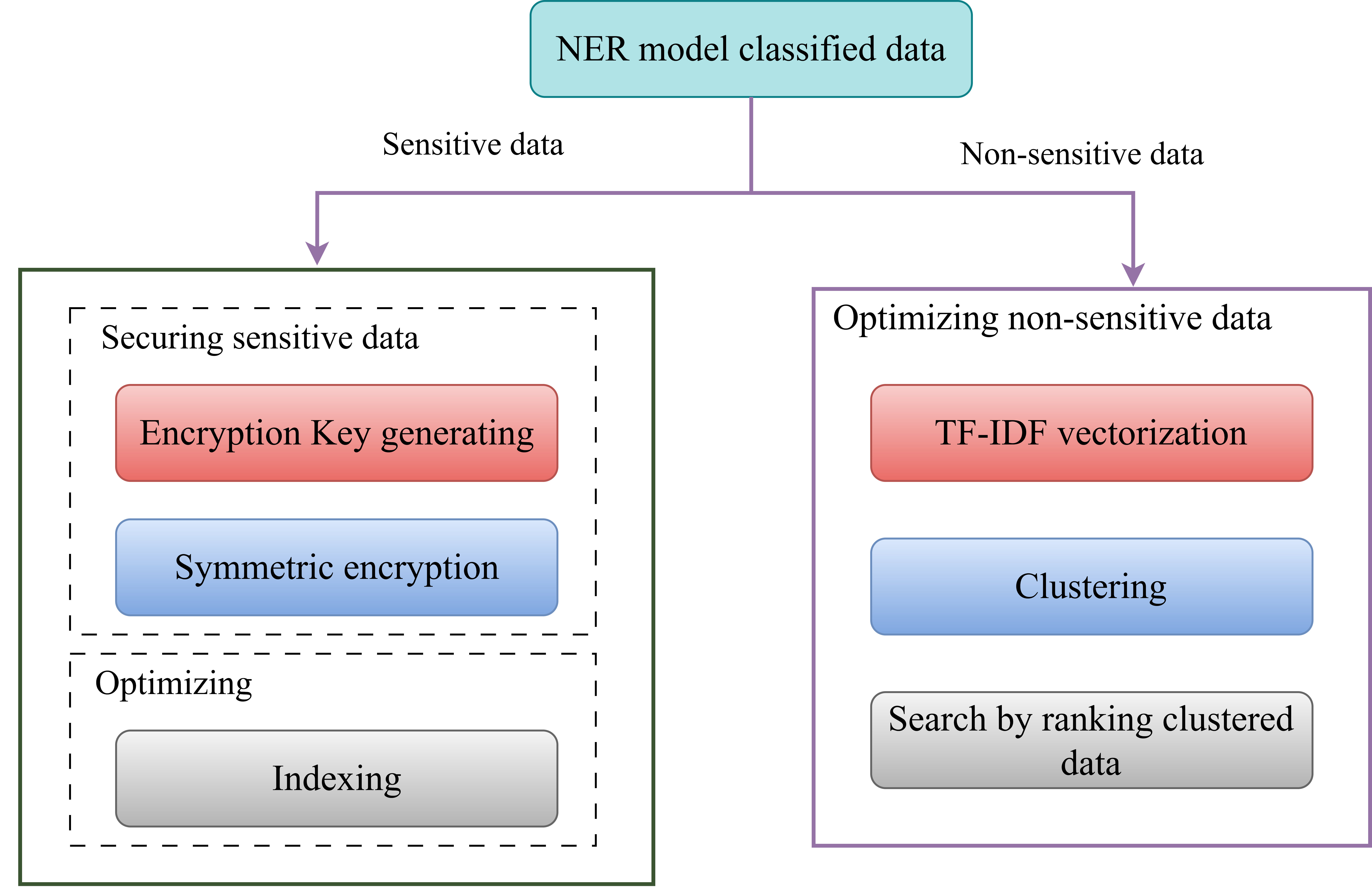}
	\caption{ Database securing and optimization process flow diagram}\label{figure:optimize_diagram}
\end{figure}

Once the NER model classifies the data, sensitive data is then encrypted by generating a key and using a blind index for efficient retrieval. In contrast, non-sensitive data is vectorized, clustered, and ranked for faster search using a cosine similarity score. A flow diagram of the process is given in Figure \ref{figure:optimize_diagram}.

\subsubsubsection{ENCRYPTING DATA BY GENERATING A KEY} 
Symmetric encryption uses a single key for both encryption and decryption, and it is a popular form of cryptography that delivers a faster, more readily deployable form of security. It is mostly used for data processing where efficiency, speed, and complexity are central \cite{r58}. AES is one of the most extensively used symmetric encryption algorithms. There are several different complexities of AES, such as 128-bit and 256-bit, that signify increasing levels of security and deliver robust safety against unauthorized access. It is efficient and usually employed in securing and protecting sensitive data \cite{r59}. After identifying sensitive information through our NER model, it is encrypted by the AES algorithm. As it is nearly impossible to crack, it provides confidentiality for sensitive data.
\subsubsubsection{INDEXING SENSITIVE DATA} 
A blind index is an indexing technique that supports search on encrypted data without revealing the actual data. Instead of indexing sensitive columns directly, a blind index uses a deterministic cryptographic hash function, often an HMAC with a secret key, to generate a fixed-length representation of the plaintext. The hashed value, known as the blind index, is stored in conjunction with the encrypted data \cite{r60}. During a search operation, the client hashes the search term using the same function and secret key to produce a corresponding blind index value, which is then used in the database query. As 
the same input always produces the same hash (on the same key), it supports exact-match searches, even though the server is never exposed to the plaintext. Blind indexing is particularly relevant in privacy-preserving contexts, where fields such as emails, phone numbers, or credit card numbers need to be encrypted while still being searchable. It optimizes sensitive information, allowing for efficient and quick queries without compromising data privacy. Therefore, blind indexing addresses this by allowing exact-match searches on encrypted data directly with a secure, deterministic hash (blind index), which helps to reduce the overhead of searching by decrypting all records, and query performance is boosted significantly. It enables databases to support low-latency searching, even over vast volumes of encrypted data. As a result, blind indexing not only enhances data privacy but also optimizes performance by combining cryptographic security with database indexing optimization \cite{r61}.
\subsubsection{OPTIMIZATION OF NON-SENSITIVE DATA} 
\begin{figure}[htbp]
	\centering
	\includegraphics[width=\textwidth]{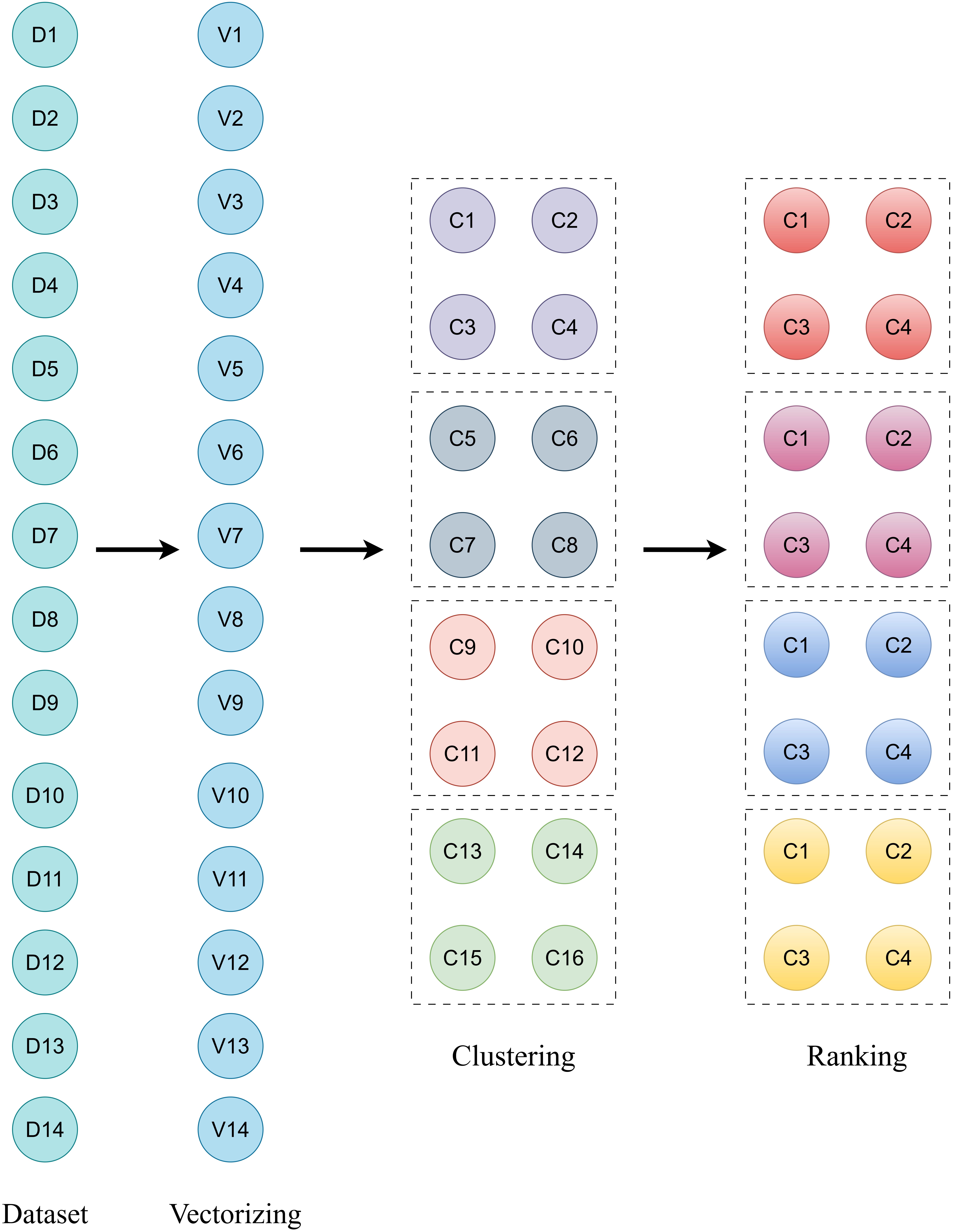}
	\caption{Clustering and ranking the database's non-sensitive information}\label{figure:cluster}
\end{figure}

\subsubsubsection{TERM FREQUENCY-INVERSE DOCUMENT FREQUENCY (TF-IDF) VECTORIZATION}
This algorithms works based on the importance of a word in a text compared to a collection of texts. It is widely used text embedding for processing text. Here, TF is depicted as term frequency, which determines how frequently a term emerges in a text \cite{r62}. It is calculated as the number of occurrences of a term divided by the total number of values in the document. Furthermore, IDF stands for inverse document frequency, which measures the rarity of a term across all texts in the corpus. It is calculated by using a logarithm function. So, TF-IDF is the product of TF and IDF, providing a higher value to those terms that are frequent in a text but rare in the entire corpus of documents. Furthermore, IDF stands for inverse document frequency, which measures the rarity of a term across all texts in the corpus. It is calculated by using a logarithm function. So, TF-IDF is the product of TF and IDF and provides a higher value to those terms that are frequent in a text but rare in the entire corpus \cite{r63}. Here, this vectorization technique is used to give high weight to those that are relevant to the context.
\subsubsubsection{CLUSTERING NON-SENSITIVE INFORMATION}
Clustering is one of the most familiar techniques for grouping similar types of data, meaning that it identifies subgroups of the data such that data points within one cluster are highly similar to each other. In contrast, data points in different clusters are highly dissimilar, thereby providing a higher value to diverse data \cite{r64} \cite{r65}. The K-means algorithm is an unsupervised algorithm that divides a dataset into N predefined, distinct subgroups. When assigning a data point, ensure that the cluster’s centroid and the squared distance between the data points are minimized. When there is less variation in the clusters, data points become more standardized \cite{r66}. For our proposed study, clustering the database based on similarities is crucial for both optimizing and preserving privacy. Our system will create clusters based on the TF-IDF-generated similarity vectors. Figure \ref{figure:cluster} demonstrates the clustering and ranking of the database. When an authorized user searches for a result, rather than searching the entire document, it first finds the cluster, as each cluster is ranked. Then, within that cluster, it searches for the intended value. In such a scenario, where the database contains 200,000 entries, retrieving a value requires searching the entire database, which is both time-consuming and inefficient. However, based on the similarities, if the cluster has only 1000 entries, then, to get the desired result, only 1000 data points need to be searched.

\subsubsubsection{Clustered data ranking by cosine similarity score}
 Cosine similarity calculates the similarity between two non-zero vectors by measuring the cosine of the angle between the vectors. In ML and data analysis, particularly in text analysis, search queries, document comparison, and recommendation systems, it is used immensely. This score is calculated by comparing the distance between data objects based on the dimensions of the feature in a dataset. A smaller distance designates a higher similarity, whereas a larger distance specifies a lower similarity. It is a metric that helps determine how similar the data objects are, regardless of their size. Measuring the similarity between sentences in a cluster data object in a dataset is considered as a vector \cite{r67}. For the query vector \(Qv\) and a document vector \(Dv\), the cosine similarity score is (1):
\begin{equation}
	\text{CosineSimilarity}(Q_v, D_v) = \frac{Q_v \cdot D_v}{\| Q_v \| \times \| D_v \|}
\end{equation}
Here, \(Qv.  Dv\): is the dot product of query and document vector, and \(|(|Qv|)|\times||Dv||\) is the Euclidean norm of the vectors.
\subsection{ALGORITHMS FOR DETECTING SENSITIVE INFORMATION}
\subsubsection{BIDIRECTIONAL LONG SHORT-TERM MEMORY NETWORKS (BILSTM)}
BiLSTM is a powerful neural network architecture that builds upon the capabilities of LSTM, a type of recurrent neural network designed for sequence modeling \cite{r68}. It consists of two LSTM layers: one processes the input from beginning to end (in the forward direction), and another processes the input from end to start (in the backward direction). Their outputs are summed up, providing the model with a better understanding of the meaning of each word by considering both its history and future.   The outputs at each time step are concatenated to produce the final output \(y_t\):
\begin{align}
	\vec{h}_t &= f(\vec{W}_x \vec{x}_t + \vec{W}_h \vec{h}_{t-1} + \vec{b}_h) \\
	\overleftarrow{h_t} &= f(\overleftarrow{W}_x x_t + \overleftarrow{W}_h \overleftarrow{h}_{t+1} + \overleftarrow{b}_h) \\
	y_t &= [\vec{h}_t, \overleftarrow{h}_t]
\end{align}

Here, $\vec{h}_t$ and $\vec{h}_{t-1}$ are the state vectors for the forward direction. 
$\overleftarrow{h_t}$ and $\overleftarrow{h}_{t+1}$ are the state vectors for the backward direction. 
$\vec{W}_x$ and $\overleftarrow{W}_x$ are the input matrices for the forward and backward direction, respectively. 
$\vec{W}_h$ and $\overleftarrow{W}_h$ are the weight matrices for the forward and backward direction, respectively. 
Finally, $\vec{b}_h$ and $\overleftarrow{b}_h$ are the biases.

BiLSTM proves useful for tasks such as NER, where having an understanding of the context surrounding a word enhances entity recognition precision. BiLSTM significantly contributes to query understanding and safe data processing. It works optimally by detecting multi-word named entities, such as physician names, organization names, and timestamps, through a full context scan of individual words. Two-way processing, as such, again supports increased understanding of semantically ambiguous or domain-specific words \cite{r69}.
\subsubsection{CONVOLUTIONAL NEURAL NETWORK (CNN)}
Deep learning algorithms, such as CNN, employs convolutional filters, known as kernels, that scan the word embeddings of the input to find meaningful local patterns, such as prefixes, suffixes, or set expressions. These yield high-level features from small chunks of text while pooling layers down-sample the dimension, enabling efficient computation to be performed \cite{r70}. CNNs are particularly good at  tasks like image classification, object detection, and segmentation because CNN can automatically learn and adaptively discover spatial hierarchies of features using backpropagation \cite{r71}. It draw on three main types of layers to learn efficiently about spatial and temporal interactions: The Convolutional Layer constitutes the heart of CNN. Each filter is designed to detect local characteristics, such as edges, textures, or shapes, by computing element-wise summations and multiplications in small spatial regions, but across the Depth of the entire input\cite{r72} \cite{r73}.
\subsubsection{DEEP BELIEF NETWORK}
DBNs are higher-level probabilistic generative models built by stacking several successive layers of Restricted Boltzmann Machines. They especially excel at hierarchical feature learning through discovering successively higher and more abstract representations from raw, unlabeled data step by step. DBNs are typically trained in two stages: the first being an unsupervised pre-training stage in which each RBM layer is learned greedily to initialize the network weights optimally in advance, and the second being a supervised fine-tuning stage to fine-tune the model to specific tasks. This two-step training enables DBNs to unscramble the actual features from noisy or uncertain inputs successfully. On sequence models, DBNs are showing improved performance on tasks like NER by facilitating richer and deeper representations that preserve the natural structure of input data, allowing the model to process meaning and context more effectively \cite{r74} \cite{r75}.
\subsubsection{BIDIRECTIONAL ENCODER REPRESENTATIONS FROM TRANSFORMERS (BERT)}
BERT is one of the newer transformer-based language models with a particular intent to learn deep word meaning knowledge by analyzing context for each word in two directions—left and right—simultaneously. Its global perspective enables the model to better capture more subtle language phenomena, long-distance word dependence \cite{r76}. The first is masked language modeling, where words in a sentence are randomly masked, and the Model is prompted to predict them based on context provided by the exposed words alone \cite{r77}.The second one, Next Sentence Prediction, is designed to train the model to predict whether two given sentences logically follow one another in sequence, thereby learning semantic relations between sentences.\cite{r78}.
\subsubsection{XLNET } 
XLNet is a transformer-based model that  left-to-right or right-to-left order. This method enables efficient management of bidirectional context without relying on masked tokens or artificial token masking for some of them \cite{r79}. XLNet adopts the Transformer-XL architecture, which features a recurrence mechanism that extends the context window beyond fixed-sized segments to manage long-range text dependencies efficiently. This makes XLNet particularly well-suited to learning and processing longer text, where it is desirable to have the ability to see how distant words relate to one another. By predicting tokens in random orders, XLNet is capable of learning patterns for higher-level dependencies and modeling language more effectively. This translates to improved performance across a wide spectrum of NLP tasks, including language modeling, text classification, and entity recognition. XLNet predicts tokens \cite{r80}
\subsection{PRIVACY ANALYSIS OF THE PROPOSED MODEL }	
According to the proposed model, privacy is ensured by encryption and a blind index for sensitive information, whereas a secure vectorized search is used for non-sensitive information. Blind index for sensitive information is a technique that allows searching over encrypted or hashed data without directly revealing any sensitive information. When a user inserts data containing sensitive information, such as an email, phone number, or credit card number, our NER system will automatically identify this information by providing Q query. Encrypt it by identifying those. For encrypting this sort of data, we have used AES algorithms.  When a data owner (DO) inserts an entry into a cloud server (CS), it sends query Q to the CS. The system (S) receives the Q, identifies all the sensitive information by our NER model that cannot be revealed to the CS, and generates a secret key by using \(K\), S encrypts the private information of the Q. Using the same K, S generates a blind index for that value. In the CS store, both Index and Q are present. Therefore, outside the S, which is considered reliable, the CS does not have any information about the Q and the K. Even if some hack the CS, they will not be able to reengineer the Q and retrieve the information. 
For the non-sensitive part, we have a vectorized form of the text. Therefore, without encrypting the normal text, the CS will not be able to gather information, as all the non-sensitive information is in number form via the TF-IDF; capturing their semantic content and term importance is a nearly arduous task.
\subsection{QUERY OPTIMIZATION  }
Encrypting data generally increases complexity during search due to the trade-off between confidentiality and query efficiency. As a result, to increase efficiency, our system can also optimize query retrieval. Blind indexing facilitates rapid searches by enabling efficient searches over encrypted data without the need for decryption, and provides quick and direct access to relevant records. For non-sensitive data retrieval, search queries are converted into V vectors, which transform the documents into numerical vectors using TF-IDF, and the main content is hidden. K-means clustering is used to determine in which cluster our desired Q is present. When the CS locates the cluster, the ranking is created specifically by the cosine similarity comparison between the search query Q and the V vectors of the documents within that cluster; the most relevant documents to the query are returned first. This combined strategy is particularly advantageous in contexts such as secure messaging, medical records, or financial logs, where the importance of both privacy and usability is paramount.

\section{Result}
This section presents the comprehensive results of our proposed system. Our experimental model is divided into two sections. The first part involves identifying sensitive data (emails, phone numbers, and credit card numbers). In contrast, the second part focuses on securing sensitive data by encrypting it, and finally, optimizing our database. This integration enables the automatic identification, encryption, and indexing of sensitive fields, thereby facilitating secure, efficient, and compliant querying. After completing the data preprocessing step, which played a significant role in producing better results, we trained our proposed model by using DL algorithms, LSTM, CNN, DBN, hybrid models, and transformer-based models. We evaluated their results in terms of accuracy, precision, recall, and F1 score. The second phase of our proposed model, which identified privacy-aware data using our NER system, was encrypted using AES encryption. On the other hand, for optimizing the database, a blind index is used for sensitive information, while non-sensitive texts are vectorized and clustered for fast and efficient retrieval. Furthermore, the integration of NER with the NoSQL database query optimization framework ensures that sensitive entities can be efficiently queried while maintaining high levels of data privacy.
\subsection{EXPERIMENTAL SETUP}
Tests and experiments were carried out on a stand-alone computer running Windows 11 (64-bit) with 16 GB of RAM. The Python programming language (version 3.13.3) was used for the implementation. Jupyter Notebook, a local development environment, was used to run all of the code. Exploratory data analysis is done with NumPy and Pandas. SpaCy is used for preparing natural language texts, including named entities, tokenization, lemmatization, word-to-vector, and part-of-speech tagging. \texttt{to\_categorical} turned text into vectors, while \texttt{pad\_sequences} is used to make all inputs the same length. Scikit-learn's \texttt{train\_test\_split} and \texttt{classification\_report} are used for splitting and producing the classification report. Whereas TensorFlow, a popular DL framework for analyzing neural networks, was employed for building and testing our DL model. For generating graphs and plots, the matplotlib and seaborn libraries are used. Hashlib is used to generate a blind index, and the AES package from the Crypto module is used. Cipher package for encrypting sensitive information. For clustering and extracting text features, K-Means and TfidfVectorizer from Scikit-learn are utilized.
\subsection{NER MODEL PERFORMANCE EVALUATION}
We applied BiLSTM, CNN algorithms, and hybrid models that combined LSTM-CNN, DBN-LSTM, and two transformer-based algorithms, BERT and XLNET, to assess the performance of our proposed system’s NER model. The dataset was split by using an 80:20 ratio, and 10-fold cross-validation was used to validate the dataset. Among these DL models, the BiLSTM parameter information includes the following: two hidden layers, a maximum sequence length of 100, a dropout rate between 0.2 and 0.5 (with the optimal result obtained at 0.2), a learning rate of 0.001, and a warmup proportion of 0.1. The batch size varies from 16 to 32 and 64; however, the best result is produced by a batch size of 32, and the number of epochs ranges from 25 to 40. The embedding dimension value is 128, the input shape is (100,), and the Adam optimizer is used. On the other hand, for these two hybrid models, the batch size is 32, the maximum sequence length is 100, and the Embedding Dimension, Epochs, and Input Shape are similar to those of BiLSTM. Nevertheless, the primary difference is in the hidden layers. For the LSTM-CNN hybrid model, the first hidden layer is a convolution layer with a filter size of 64, several kernels of 3, and the same padding. LSTM is the second layer with a filter size of 64, whereas the third layer is a dense layer with a value of 128, and ReLU is the activation function. For our proposed model’s second hybrid model, merging LSTM and DBN with recurrent units of 34, a convolution layer with a filter size of 64 and a kernel size of 3 is the second layer, and the third layer is an LSTM layer with a size of 128. The hyperparameter list for training DL algorithms is given in Table \ref{tab:model_architectures}. For the transformer-based models, we also used various parameters to get the best output. BERT and XLNet were fine-tuned using a maximum sequence length of 128, a batch size of 32, and a learning rate of 2e-5. Both models were trained for 30 epochs using the Adam optimizer, with early stopping applied to prevent overfitting.
\begin{table}[tbp]  
	\centering
	\caption{Hyperparameter values of the DL algorithms}
	\label{tab:model_architectures}
	\begin{tabular}{p{0.15\linewidth} | p{0.25\linewidth} | p{0.25\linewidth} | p{0.25\linewidth}}
		\toprule
		\textbf{Category} & \textbf{BiLSTM model} & \textbf{CNN + LSTM model} & \textbf{DBM + LSTM model} \\
		\hline
		\midrule
		Batch Size & 16, 32, 64 & 32 & 32 \\
		\hline
		Max Sequence Len & 100 & 100 & 100 \\
		\hline
		Layer 1 & Bidirectional LSTM (64) & Conv1D (64, kernel=3, same) & Dense layer (128) \\
		\hline
		Layer 2 & TimeDistributed Dense (64, relu) & LSTM (64) & Conv1D (64, kernel=3) \\
		\hline
		Layer 3 & N/A & Dense (128, relu) & LSTM (128) \\
		\hline
		Embedding Dim & 128 & 128 & 128 \\
		\hline
		Optimizer & adam & adam & adam \\
		\hline
		Epochs & 30 & 25 & 40 \\
		\hline
		Label Format & padded, int, expand\_dims(y, -1) & padded, int, no expand\_dims & padded, int, no expand\_dims \\
		\hline
		Input Shape & (100,) & (100,) & (100,) \\
		\bottomrule
	\end{tabular}
\end{table}
\subsubsection{PERFORMANCE EVALUATION BY BILSTM}
\begin{figure}[htbp]
	\centering
	\includegraphics[width=\textwidth]{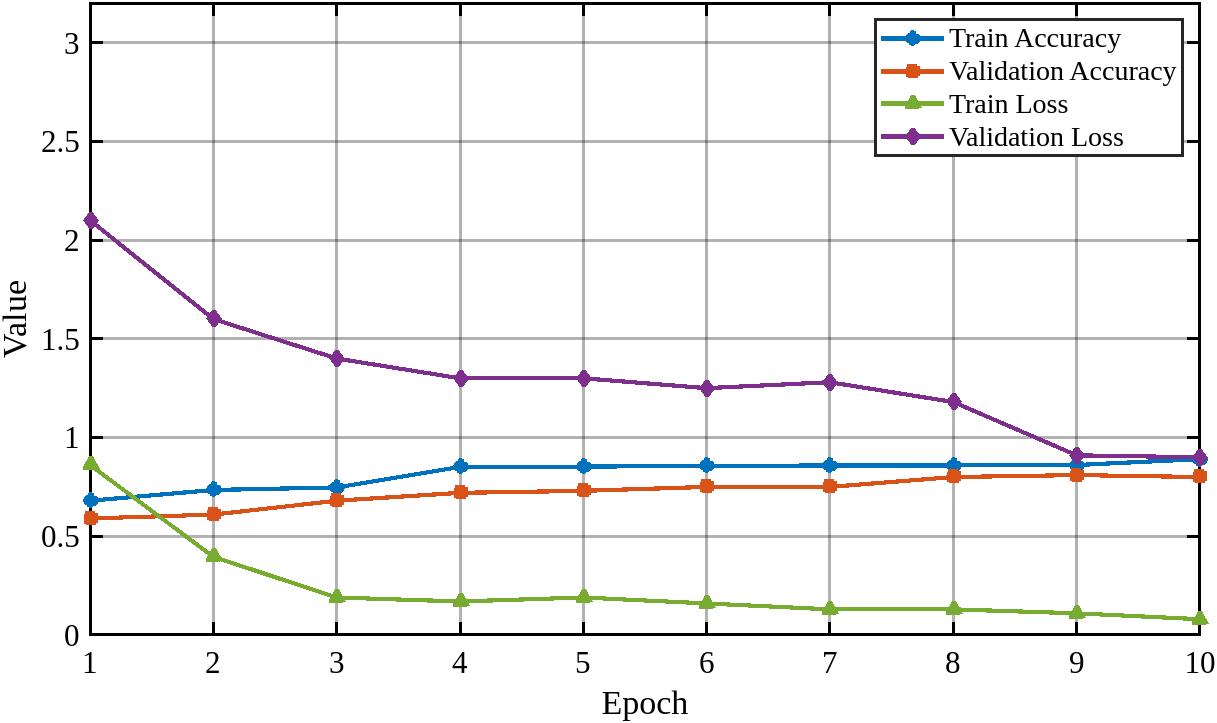}
	\caption{Result of four metrics by using the BiLSTM algorithm for 10 epochs.}\label{figure:BiLSTM_epoch_performance}
\end{figure}

The results of the BiLSTM algorithms for the first 10 epochs out of 25 epochs are presented in Table \ref{tab:BILSTM_epochs}. From the table, Training accuracy reaches 85\% after 4 epochs, whereas it was 68\% in the first epoch. Training accuracy increased steadily after epoch number 3. On the other hand, training loss is 0.861 in the first epoch and shows a sharp decline in the second epoch (0.396) and continuously declines in every iteration. After 10 iterations, the training loss approaches zero, whereas the validation accuracy starts at 59\% in the first epoch. However, slight fluctuations occur; still, it remains consistently above 80\%. Furthermore, the validation loss starts at 2.1 and steadily decreases to 0.9 by epoch 10, as the model gradually improves on the training data, while the validation error slowly rises. The NER model exhibits outstanding performance using the BiLSTM algorithm (Table \ref{tab:BILSTM_entity_performance}) across all sensitive entities, including \texttt{CREDIT\_CARD}, URL, PHONE, MONEY, SSN, and the non-entity class O, achieving precision, recall, and F1-scores above 85\%, demonstrating perfect identification of these tokens. The only exception is the email, where the model achieved a precision of 83\%, a recall of 77\%, and an F1 score of 82\% Figure \ref{figure:BiLSTM_epoch_performance}. Therefore, the results depict that the model is highly precise and robust for identifying sensitive information, making it appropriate for tasks including privacy-aware data processing or secure information extraction.
\begin{table}[tbp]  
	\centering
	\caption{BiLSTM algorithms' results for the first 10 epochs}
	\label{tab:BILSTM_epochs}
	\begin{tabular}{>{\centering\arraybackslash}p{0.16\linewidth} | >{\centering\arraybackslash}p{0.16\linewidth} | >{\centering\arraybackslash}p{0.16\linewidth} | >{\centering\arraybackslash}p{0.16\linewidth} | >{\centering\arraybackslash}p{0.16\linewidth}}
		\toprule
		\textbf{Epoch} & \textbf{Train Accuracy} & \textbf{Train Loss} & \textbf{Validation Accuracy} & \textbf{Validation Loss} \\
		\hline
		\midrule
		1 & 0.68 & 0.861 & 0.59 & 2.1 \\
		\hline
		2 & 0.7345 & 0.396 & 0.61 & 1.6 \\
		\hline
		3 & 0.7474 & 0.19 & 0.68 & 1.4 \\
		\hline
		4 & 0.8511 & 0.17 & 0.72 & 1.3 \\
		\hline
		5 & 0.8528 & 0.19 & 0.73 & 1.3 \\
		\hline
		6 & 0.8562 & 0.16 & 0.75 & 1.25 \\
		\hline
		7 & 0.8586 & 0.13 & 0.75 & 1.28 \\
		\hline
		8 & 0.8589 & 0.13 & 0.8 & 1.18 \\
		\hline
		9 & 0.8607 & 0.11 & 0.81 & 0.91 \\
		\hline
		10 & 0.8908 & 0.08 & 0.8 & 0.9 \\
		\bottomrule
	\end{tabular}
\end{table}

\begin{table}[tbp]  
	\centering
	\caption{Entity performance metrics of BiLSTM algorithm}
	\label{tab:BILSTM_entity_performance}
	\begin{tabular}{>{\centering\arraybackslash}p{0.23\linewidth} | >{\centering\arraybackslash}p{0.23\linewidth} | >{\centering\arraybackslash}p{0.23\linewidth} | >{\centering\arraybackslash}p{0.23\linewidth}}
		\toprule
		\textbf{Entity} & \textbf{Precision (\%)} & \textbf{Recall (\%)} & \textbf{F1-Score (\%)} \\
		\hline
		\midrule
		B-CREDIT\_CARD & 93 & 91 & 94 \\
		\hline
		B-EMAIL & 83 & 77 & 82 \\
		\hline
		B-MONEY & 96 & 88 & 90 \\
		\hline
		B-PHONE & 87 & 89 & 83 \\
		\hline
		B-SSN & 89 & 86 & 79 \\
		\hline
		B-URL & 91 & 87 & 96 \\
		\hline
		O & 94 & 97 & 92 \\
		\bottomrule
	\end{tabular}
\end{table}

\subsubsection{PERFORMANCE EVALUATION BY CNN-LSTM}
\begin{figure}[htbp]
	\centering
	\includegraphics[width=\textwidth]{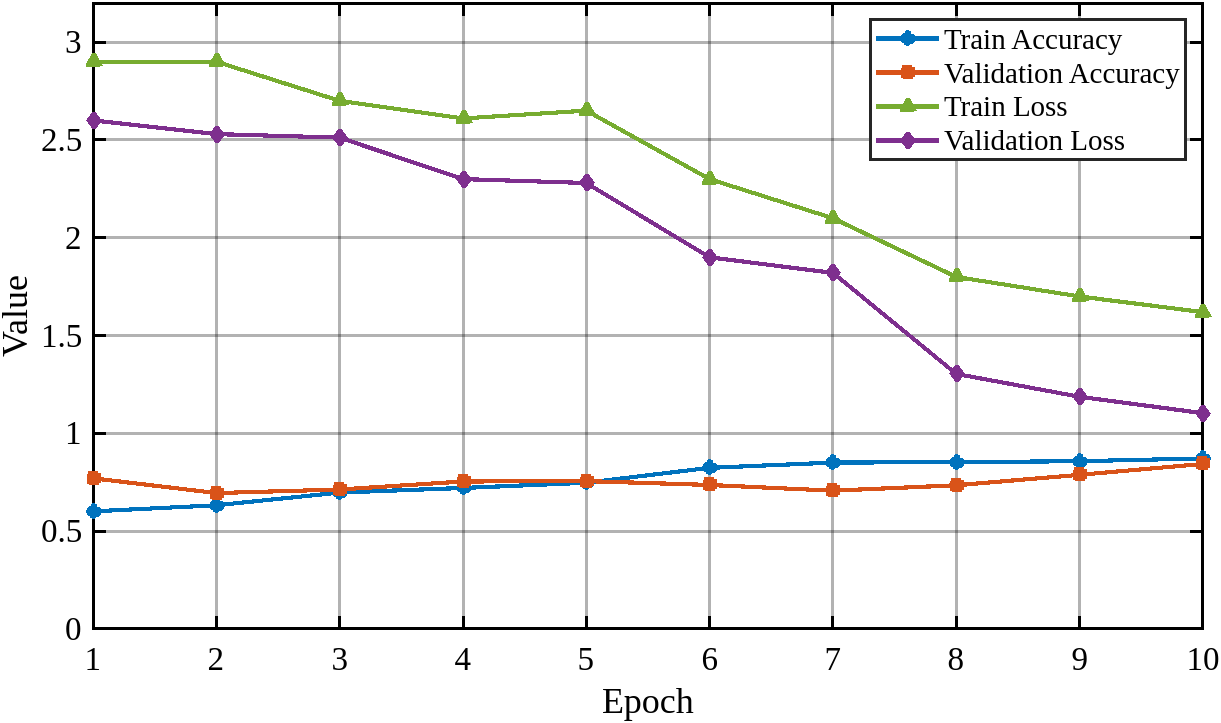}
	\caption{ Result of four metrics by using the CNN-LSTM algorithm for 10 epochs}\label{figure:CNN_LSTM_epoch_performance}
\end{figure}

Our proposed hybrid model, utilizing CNN-LSTM, exhibits a strong learning trend for the first 10 epochs, as illustrated in Table \ref{tab:cnn_lstm_epoch}. In the first epoch, training accuracy is 61.15\% and the loss is 2.9. The accuracy was 87.2\% with a significantly reduced loss of 1.62 after 10 epochs. On the other hand, validation performance demonstrates more fluctuation, with an accuracy of 77.05\% in the first epoch, but astoundingly drops to 69.5\% in the next epoch, then gradually increases, accompanied by a rise in loss to 1.8206, as shown in Figure \ref{figure:CNN_LSTM_epoch_performance}. From Table \ref{tab:CNN_lstm_entity_performance}, the model's performance on different entities varied; among them, the \texttt{CREDIT\_CARD} entity has the highest result, with a precision of 0.88, a recall of 0.84, and an F1-score of 0.87.  Additionally, PHONE and MONEY also performed well, with F1-scores of 0.87 and 0.82, respectively. B-Money recall was 0.1 higher than precision. Whilst two entities, EMAIL and URL, show the weakest results, with F1-scores of 0.72 and 0.67, respectively. The non-entity class accomplished a strong F1 score of 0.83.
\begin{table}[tbp]  
	\centering
	\caption{CNN-LSTM algorithm results for the first 10 epochs}
	\label{tab:cnn_lstm_epoch}
	\begin{tabular}{>{\centering\arraybackslash}p{0.14\linewidth} | >{\centering\arraybackslash}p{0.18\linewidth} | >{\centering\arraybackslash}p{0.18\linewidth} | >{\centering\arraybackslash}p{0.18\linewidth} | >{\centering\arraybackslash}p{0.18\linewidth}}
		\toprule
		\textbf{Epoch} & \textbf{Train Accuracy} & \textbf{Validation Accuracy} & \textbf{Train Loss} & \textbf{Validation Loss} \\
		\hline
		\midrule
		1 & 0.6015 & 0.7705 & 2.9 & 2.6 \\
		\hline
		2 & 0.6328 & 0.695 & 2.9 & 2.53 \\
		\hline
		3 & 0.6977 & 0.7133 & 2.7 & 2.512 \\
		\hline
		4 & 0.721 & 0.7542 & 2.61 & 2.3 \\
		\hline
		5 & 0.748 & 0.7562 & 2.65 & 2.28 \\
		\hline
		6 & 0.824 & 0.7365 & 2.3 & 1.90 \\
		\hline
		7 & 0.851 & 0.7072 & 2.1 & 1.8206 \\
		\hline
		8 & 0.853 & 0.7348 & 1.8 & 1.3042 \\
		\hline
		9 & 0.858 & 0.7884 & 1.7 & 1.1872 \\
		\hline
		10 & 0.872 & 0.8439 & 1.62 & 1.1035 \\
		\bottomrule
	\end{tabular}
\end{table}

\begin{table}[tbp]  
	\centering
	\caption{Entity performance metrics of CNN-LSTM algorithm}
	\label{tab:CNN_lstm_entity_performance}
	\begin{tabular}{p{0.25\linewidth} | p{0.25\linewidth} | p{0.25\linewidth} | p{0.25\linewidth}}
		\toprule
		\textbf{Entity} & \textbf{Precision} & \textbf{Recall} & \textbf{F1-Score} \\
		\midrule
		B-CREDIT\_CARD & 0.88 & 0.84 & 0.87 \\
		\hline
		B-EMAIL & 0.63 & 0.67 & 0.72 \\
		\hline
		B-MONEY & 0.86 & 0.88 & 0.87 \\
		\hline
		B-PHONE & 0.79 & 0.68 & 0.73 \\
		\hline
		B-SSN & 0.60 & 0.68 & 0.67 \\
		\hline
		B-URL & 0.79 & 0.80 & 0.82 \\
		\hline
		O & 0.87 & 0.91 & 0.83 \\
		\bottomrule
	\end{tabular}
\end{table}

\subsubsection{PERFORMANCE EVALUATION BY DBN-LSTM}
\begin{figure}[htbp]
	\centering
	\includegraphics[width=\textwidth]{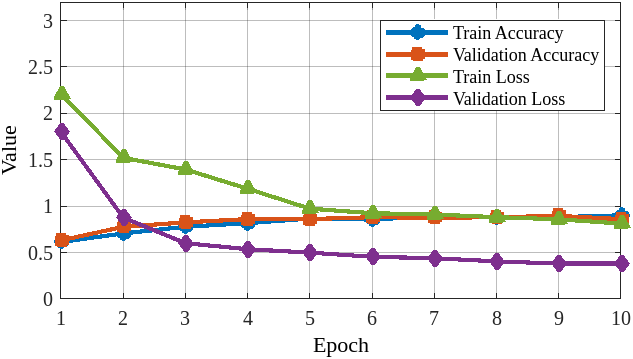}
	\caption{ Result of four metrics by using the DBN-LSTM algorithm for 10 epochs}\label{figure:dbn_lstm_epoch_performance}
\end{figure}

The hybrid model DBN-LSTM's performance is outstanding in almost all entities, particularly in learning and generalization capabilities, as shown in Table \ref{tab:DBN_lstm_epoch} throughout the first 10 training epochs out of 30 epochs. Originally, the training accuracy is comparatively low at 61.48\%, with a corresponding loss of 2.2, and then increases continuously (Figure \ref{figure:dbn_lstm_epoch_performance}). By the 10th epoch, the model’s validation performance reached 86.11\%. Notably, the validation loss shows a significant drop from epoch 2 onwards, falling to 0.38 by the final epoch. The evaluation results from different entities in Table \ref{tab:DBN_LSTM_Entity} highlight notable variations in the model's performance. Entities with distinct and structured formats, such as \texttt{CREDIT\_CARD} and PHONE, achieved the highest performance, with F1-scores of 0.96 and 0.89, respectively. Likewise, MONEY's F1-score of 0.75 and SSN's F1-score of 0.74 prove satisfactory results. On the other hand, with an F1-score of 0.68, EMAIL has a relatively low Recall of 0.64, indicating the model struggles to identify all relevant email entities. Finally, the O label, denoting non-entity tokens, maintains a strong F1 score of 0.79, reflecting consistent performance in distinguishing between entity and non-entity text.
\begin{table}[tbp]  
	\centering
	\caption{DBN-LSTM algorithm results for the first 10 epochs}
	\label{tab:DBN_lstm_epoch}
	\begin{tabular}{p{0.15\linewidth} | p{0.2\linewidth} | p{0.2\linewidth} | p{0.2\linewidth} | p{0.2\linewidth}}
		\toprule
		\textbf{Epoch} & \textbf{Train Accuracy} & \textbf{Train Loss} & \textbf{Validation Accuracy} & \textbf{Validation Loss} \\
		\hline
		\midrule
		1 & 0.6148 & 2.2086 & 0.6319 & 1.8002 \\
		\hline
		2 & 0.7066 & 1.5223 & 0.7778 & 0.8781 \\
		\hline
		3 & 0.7785 & 1.3961 & 0.8264 & 0.6015 \\
		\hline
		4 & 0.8171 & 1.1881 & 0.8611 & 0.5367 \\
		\hline
		5 & 0.8670 & 0.9737 & 0.8611 & 0.4999 \\
		\hline
		6 & 0.8566 & 0.92445 & 0.8819 & 0.4554 \\
		\hline
		7 & 0.8927 & 0.9107 & 0.8750 & 0.4398 \\
		\hline
		8 & 0.8802 & 0.882 & 0.8819 & 0.4058 \\
		\hline
		9 & 0.8822 & 0.859 & 0.8958 & 0.3817 \\
		\hline
		10 & 0.9019 & 0.8133 & 0.8611 & 0.3837 \\
		\bottomrule
	\end{tabular}
\end{table}
\begin{table}[tbp]  
	\centering
	\caption{Entity performance metrics of the DBN-LSTM algorithm}
	\label{tab:DBN_LSTM_Entity}
	\begin{tabular}{p{0.25\linewidth} | p{0.25\linewidth} | p{0.25\linewidth} | p{0.25\linewidth}}
		\toprule
		\textbf{Entity} & \textbf{Precision} & \textbf{Recall} & \textbf{F1-Score} \\
		\midrule
		B-CREDIT\_CARD & 0.95 & 0.98 & 0.96 \\
		\hline
		B-EMAIL & 0.72 & 0.64 & 0.68 \\
		\hline
		B-MONEY & 0.91 & 0.87 & 0.89 \\
		\hline
		B-PHONE & 0.84 & 0.66 & 0.74 \\
		\hline
		B-SSN & 0.62 & 0.63 & 0.62 \\
		\hline
		B-URL & 0.73 & 0.77 & 0.75 \\
		\hline
		O & 0.76 & 0.82 & 0.79 \\
		\bottomrule
	\end{tabular}
\end{table}

\subsubsection{ALL ALGORITHMS’ FINAL RESULTS}
\begin{figure}[htbp]
	\centering
	\includegraphics[width=\textwidth]{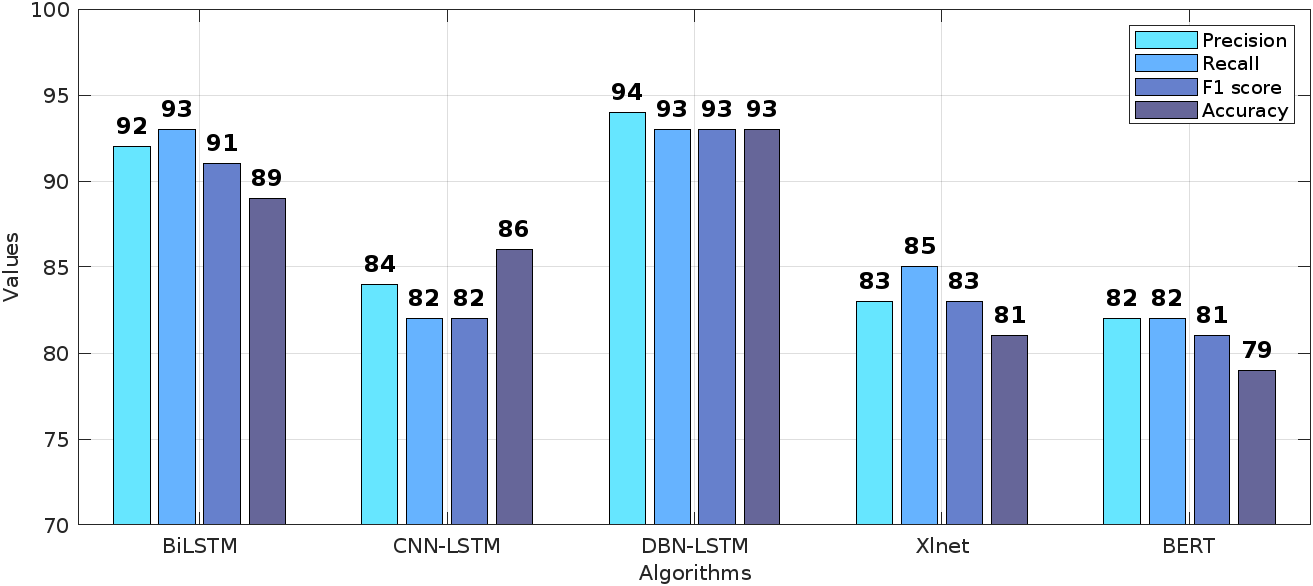}
	\caption{Performance comparison of five models for detecting entities from the schema by the NER model}\label{figure:all_algorithm_performance}
\end{figure}

To change the default, adjust the template as follows. Table \ref{tab:algorithm_comparison} compares several DL algorithms and two pre-trained transformer models, such as BERT and XLNET, final results after completing the training. DBN with LSTM is the top performer with the highest accuracy (93\%), precision (94\%), recall (93\%), and F1-score (93\%). The LSTM model also performs well, especially in recall and precision (93\% and 92\%, respectively). In contrast, CNN, although slightly higher in accuracy value, which is 90\%, falls behind in precision, and F1 score, which is 2\% , and 10\% lower, respectively and the hybrid model CNN-LSTM has balanced performance (accuracy 86\%, F1 score 82\%), also depicted in Figure \ref{figure:all_algorithm_performance}. Transformer-based models like XLNet and BERT offer less accuracy compared to other DL models. XLNet slightly outperforms BERT in all metrics, particularly F1-score (83.4\% vs. 81\%). BERT has the lowest accuracy, 79\%, among the higher-end models.
\begin{table}[tbp]  
	\centering
	\caption{Performance Comparison of Different Algorithms}
	\label{tab:algorithm_comparison}
	\begin{tabular}{>{\centering\arraybackslash}p{0.22\linewidth} | >{\centering\arraybackslash}p{0.16\linewidth} | >{\centering\arraybackslash}p{0.16\linewidth} | >{\centering\arraybackslash}p{0.16\linewidth} | >{\centering\arraybackslash}p{0.16\linewidth}}
		\toprule
		\textbf{Algorithm} & \textbf{Accuracy (\%)} & \textbf{Precision (\%)} & \textbf{Recall (\%)} & \textbf{F1-Score (\%)} \\
		\midrule
		BiLSTM & 89 & 92 & 93 & 91 \\
		\hline
		CNN-LSTM & 86 & 84 & 82 & 82 \\
		\hline
		DBN-LSTM & 93 & 94 & 93 & 93 \\
		\hline
		Xlnet & 81 & 83 & 85 & 83.4 \\
		\hline
		BERT & 79 & 82 & 82 & 81 \\
		\hline
		BiLSTM & 89 & 92 & 93 & 91 \\
		\hline
		CNN-LSTM & 86 & 84 & 82 & 82 \\
		\bottomrule
	\end{tabular}
\end{table}

\subsection{DATABASE OPTIMIZATION PERFORMANCE EVALUATION} 
The proposed system demonstrates notable improvements in database optimization performance, with encrypted queries accelerated by blind indexing and non-sensitive data retrieval enhanced through efficient clustering and a cosine similarity score for ranked search. Additionally, it proposes efficient optimization techniques for both sensitive and non-sensitive information.
\subsubsection{SENSITIVE INFORMATION OPTIMIZATION}	
After using 16 bytes for symmetric encryption and 32 bytes for the blind index, we have verified our system by searching for sensitive information both directly and with indexing. 

\begin{figure}[htbp]
	\centering
	\includegraphics[width=\textwidth]{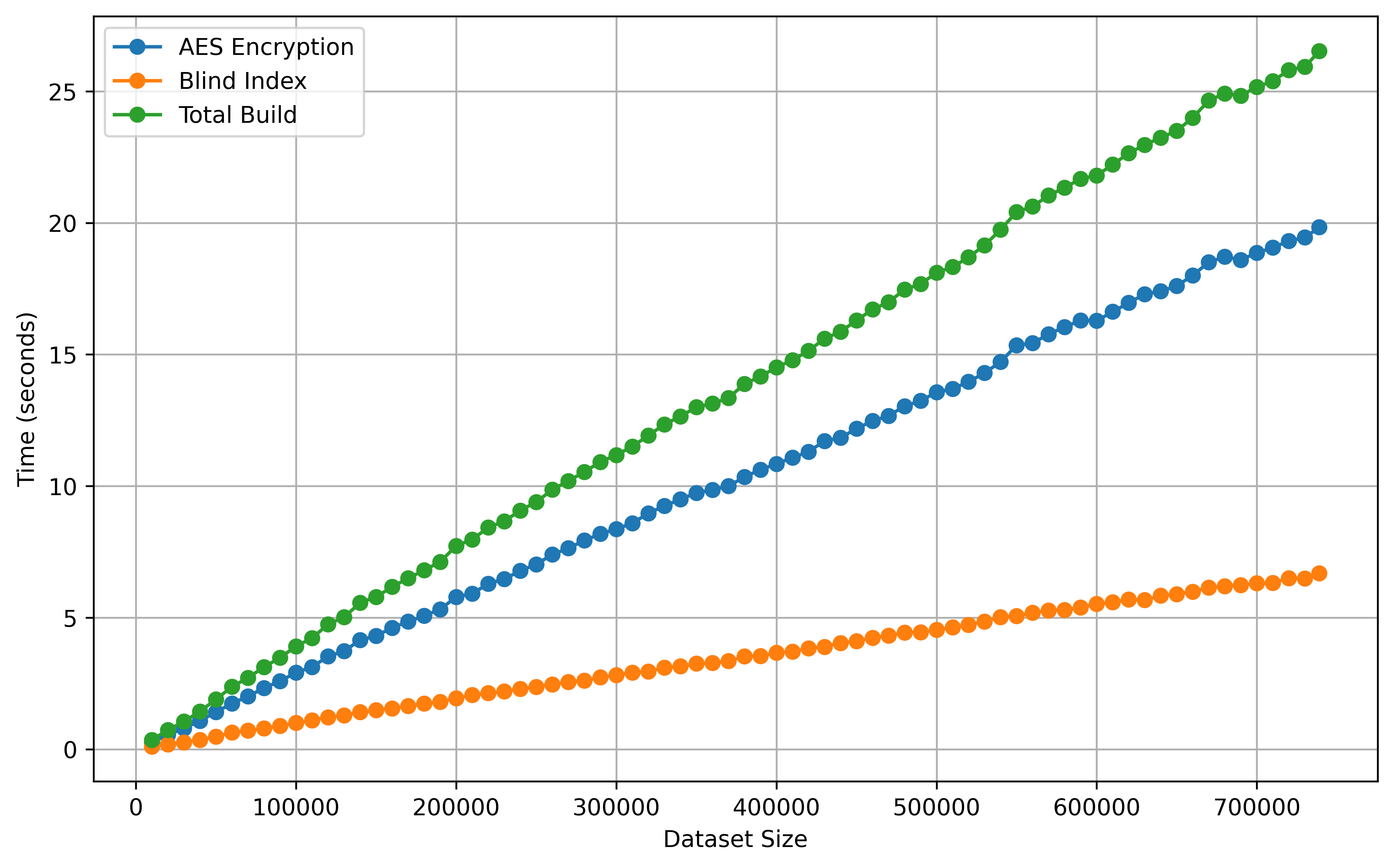}
	\caption{ Build Time vs Dataset Size (AES, Blind Index, Total Build)}\label{figure:build time}
\end{figure}

Figure \ref{figure:build time} shows how encryption time (AES), blind index generation time (HMAC),
and their total combined cost scale with dataset size. AES encryption takes longer
due to heavier block cipher operations, while HMAC hashing remains faster.
The total build cost grows linearly, showing predictable scalability.

\begin{figure}[htbp]
	\centering
	\includegraphics[width=\textwidth]{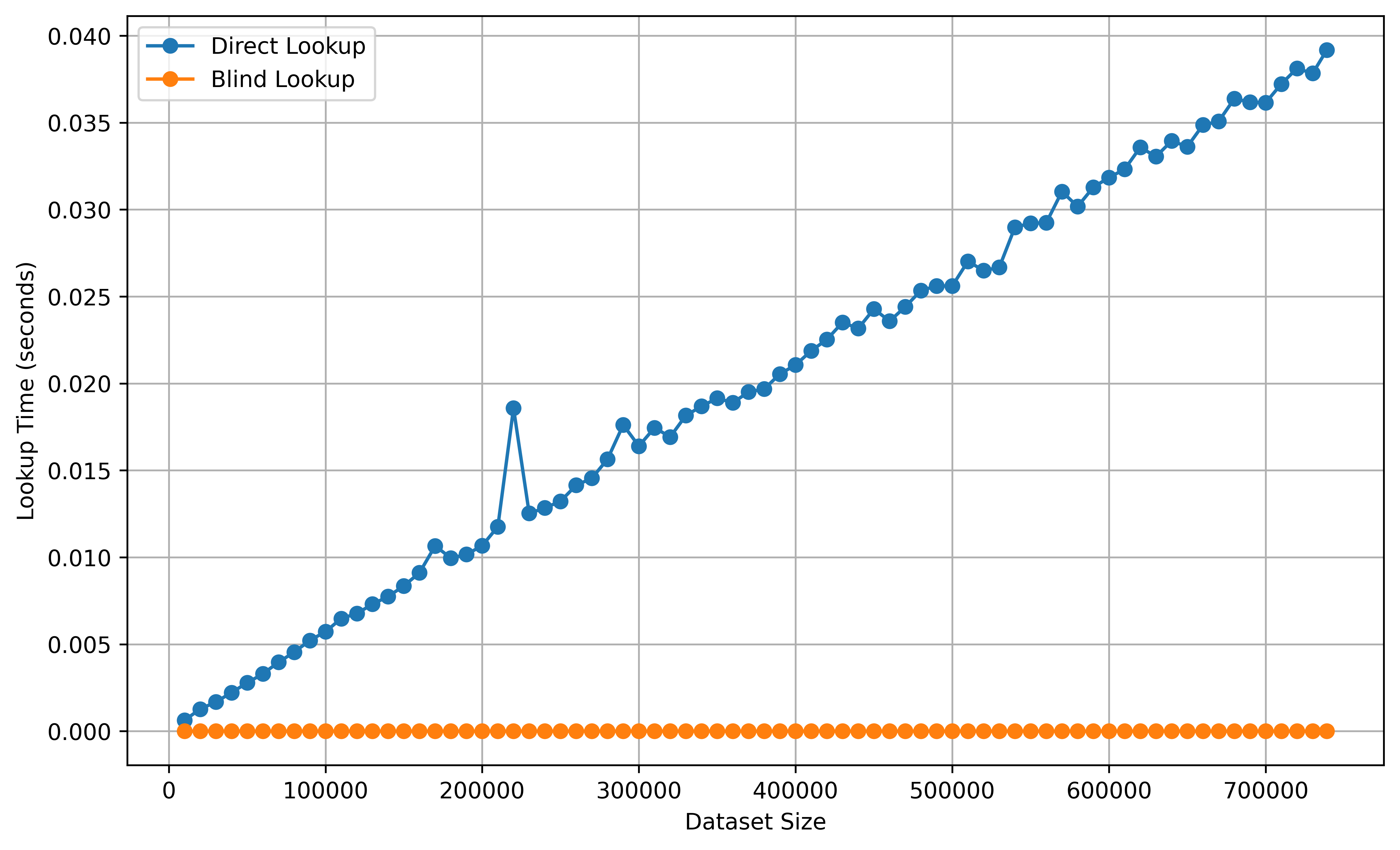}
	\caption{ Direct vs Blind Lookup time of the dataset}\label{figure:lookup_time}
\end{figure}

Figure \ref{figure:lookup_time} compares two search strategies. The direct lookup curve shows the time needed to scan plaintext data within the dataset, which performs a linear search. As the dataset grows, the search time slowly increases, indicating typical list-based lookup behavior. Conversely, the blind lookup curve (using precomputed HMAC values stored in a set) remains nearly constant regardless of dataset size. This is because blind indexing ues python set data structure which complexity is O(1) .The gap between the two lines highlights the performance advantage of blind indexing, especially for large-scale datasets where traditional plaintext scanning becomes increasingly inefficient.

\subsubsection{NON-SENSITIVE INFORMATION OPTIMIZATION}

\begin{figure}[htbp]
	\centering
	\includegraphics[width=\textwidth]{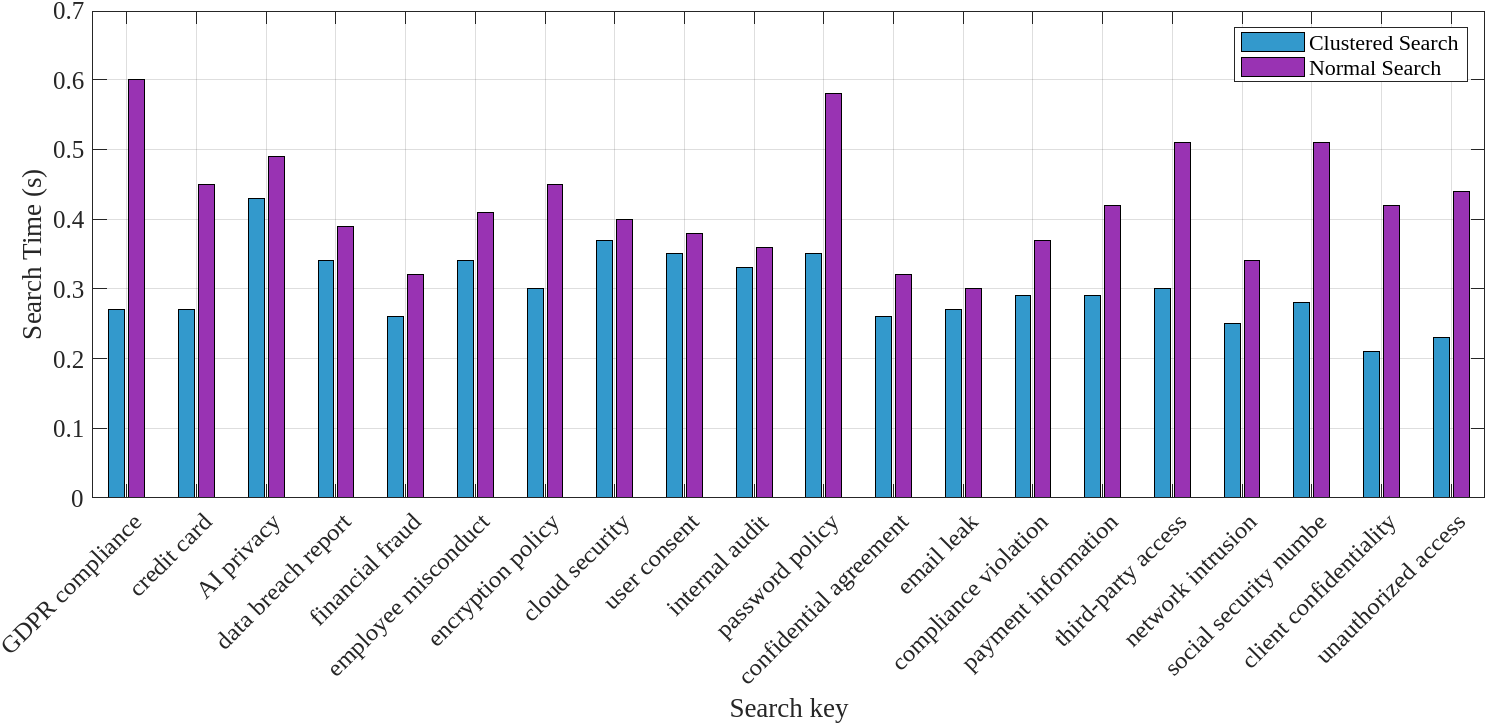}
	\caption{Non-sensitive information comparison with cluster search and normal search}\label{figure:clustered_sensitive_nonsensitive_performance}
\end{figure}

\begin{figure}[htbp]
	\centering
	\includegraphics[width=\textwidth]{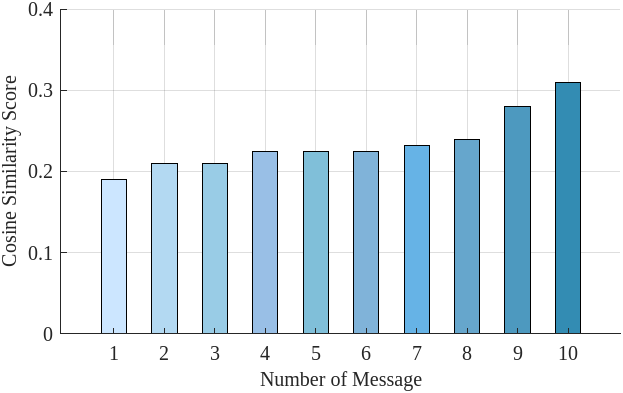}
	\caption{ Result of four metrics by using the DBN-LSTM algorithm for 10 epochs}\label{figure:up_gdpr_performance}
\end{figure}

\begin{figure}[htbp]
	\centering
	\includegraphics[width=\textwidth]{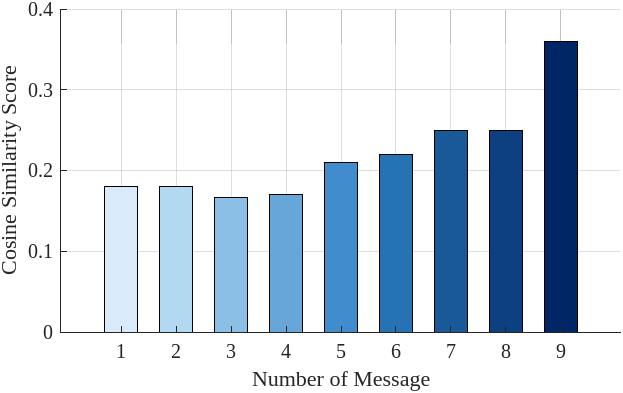}
	\caption{ Result of four metrics by using the DBN-LSTM algorithm for 10 epochs}\label{figure:up_consent_performance}
\end{figure}

\begin{figure}[htbp]
	\centering
	\includegraphics[width=\textwidth]{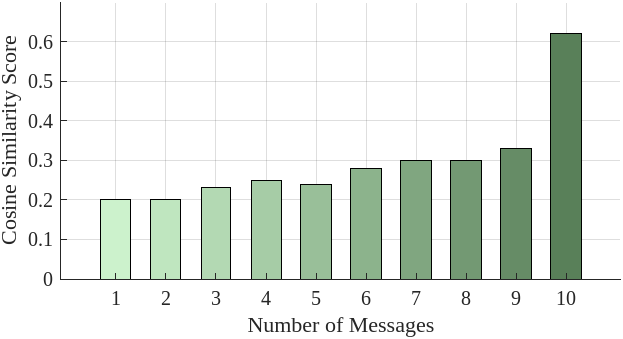}
	\caption{ Result of four metrics by using the DBN-LSTM algorithm for 10 epochs}\label{figure:up_credit_performance}
\end{figure}

Figure \ref{figure:clustered_sensitive_nonsensitive_performance} compares the search time in clustered search vs. normal search across a range of normal texts. In general, the Normal Search method takes more time across most terms, suggesting that it requires more time to retrieve non-sensitive information that is in vector form. GDPR compliance scores 0.60 in normal search, compared to 0.27 in clustered search, which is less than half the time. The password policy also shows a significant gap (0.58 vs. 0.35). For every search query, clustered search took less time, whereas normal search took a significant amount of time. Also, cosine similarity scores in a cluster for query strings such as "credit card," "GDPR," and "user consent" are shown in Figure \ref{figure:up_consent_performance}, Figure \ref{figure:up_credit_performance}, and Figure \ref{figure:up_gdpr_performance} for up to 10 matches, and our system returns the result with the highest similarity score, which produces a fast response and an accurate result.

\section{DISCUSSION}
Comparative analysis of various models against multiple performance metrics provides insights into their performance. The BiLSTM model demonstrated consistent learning and excellent generalization. CNN-LSTM applies convolutional layers to extract local spatial features and LSTM layers to extract temporal features. This combined model than CNN performed better as only CNN model lack of stability, perhaps due to the challenge of balancing feature abstraction and temporal learning. Comparatively, the highest-scoring model was DBN-LSTM with 93\% accuracy, 94\% precision, and a 93\% F1 score. This multi-stage pre-training approach adds depth to learning and regularization, allowing the model to learn more abstract, noise-robust features, which contributes to its improved performance on all measures. XLNet performed satisfactorily. However, its performance may be 
owing to a lack of fine-tuning or a lack of alignment between pretraining tasks and the domain-specific organization of the dataset. BERT, another transformer model, had the worst accuracy and relatively lower F1-score in this study. Despite its global attention mechanism, BERT underperforms due to insufficient domain-specific pretraining and a relatively smaller dataset size, thus failing to generalize effectively. From Table \ref{tab:comparison_sota}, our proposed model performed seamlessly.
\begin{table}[tbp]  
	\centering
	\caption{Comparison with State-of-the-Art Methods}
	\label{tab:comparison_sota}
	\begin{tabular}{>{\centering\arraybackslash}p{0.22\linewidth} | >{\centering\arraybackslash}p{0.14\linewidth} | >{\centering\arraybackslash}p{0.14\linewidth} | >{\centering\arraybackslash}p{0.14\linewidth} | >{\centering\arraybackslash}p{0.14\linewidth}}
		\toprule
		\textbf{Paper} & \textbf{Accuracy (\%)} & \textbf{Precision (\%)} & \textbf{Recall (\%)} & \textbf{F1-score (\%)} \\
		\hline
		\midrule
		Wilkho et al. 2025 \cite{r37} & N/A & 93.60 & 91.74 & 92.66 \\
		\hline
		Gao et al. \cite{r38} 2025 & N/A & 85.33 & 86.56 & 85.94 \\
		\hline
		H. Cho et al. 2019 \cite{r39} & N/A & 96.05 & 95.55 & 96.29 \\
		\hline
		Zhou et al. \cite{r40} 2025 & 89.09 & 89.09 & 89.49 & 89.18 \\
		\hline
		Serere et al. \cite{r41} 2025 & N/A & 91 & 90 & 90 \\
		\hline
		G. et al \cite{r42} 2025 & N/A & N/A & N/A & 87.2 \\
		\hline
		Wu et al. 2018 \cite{r43} & 72 & 75.4 & 73.3 & 74.3 \\
		\hline
		Zhang et al. \cite{r44} 2025 & N/A & 97.36 & 75.22 & 84.87 \\
		\hline
		Misra et al. \cite{r47} & 89.4 & 94.7 & 89.4 & 91.1\\
		\hline
		Our proposed model & 93.2 & 94 & 93 & 93 \\
		\bottomrule
	\end{tabular}
\end{table}
Our proposed query optimization model took significantly less time to retrieve encrypted values, particularly when checking email addresses. Time to find an email using HMAC hash in a set is O(1). The divergence demonstrates the massive scalability advantage of blind indexing. Direct search at 100K records: ~5-20 milliseconds (increases with size). Blind index search ~0.0001 milliseconds (constant, microsecond range). Therefore, ~50,000-200,000x faster lookups with blind indexing. Moreover, we also optimized non-sensitive information via TF-IDF vectorization, clustering, and search. Non-sensitive information could be retrieved faster with these mechanisms than with a standard search. To validate the proposed privacy-aware NoSQL query optimization models, the Enron Email dataset was chosen over the real NoSQL dataset due to several practical and technical considerations. First, the NoSQL database containing large amounts of sensitive data is often confidential, proprietary, or inaccessible due to data security policies and privacy regulations. Even when available, such a dataset requires expensive membership, legal permits, or a specialized environment to operate advanced NoSQL features, making them prohibitively expensive to explore in academic settings. In contrast, the Enron dataset \cite{r84} is freely available, comprising a large amount of semi-structured and unstructured data (~700,000 emails), and contains realistic sensitive information (e.g., name, email, date), which is relevant to the NoSQL environment. Previous works, such as \cite{r33} \cite{r82} \cite{r83}, have successfully utilized the Enron Corpus to validate the privacy-aware and optimization system, further justifying its use as a standard benchmark in the field of secure and optimized data access. 

By leveraging the use of DL-based NER for sensitive information, encryption, and blind indexing, along with handling non-sensitive information for fast retrieval, this research offers a scalability- and privacy-conscious NoSQL query optimization approach that enables secure and efficient access to sensitive data. The combination of cutting-edge neural models, such as DBN-LSTM, with privacy-protecting encrypted search protocols leads to enhanced accuracy as well as faster retrieval speeds, allowing NoSQL systems to be safer and smarter. The proposed privacy-aware NoSQL query optimization framework offers multi-level benefits to all types of users. The proposed system simplifies secure data integration by automating the identification of sensitive information and optimizing it for developers. For organizations and service providers, it enhances query performance and reduces the chances of data leakages. It provides user with improved privacy and data sovereignty, ensuring their sensitive and personal data are handled with maximum security and ethical responsibility.

\section{CONCLUSION} 

In this paper, we propose privacy-preserving mimic models for optimizing databases by identifying sensitive information. The novelty lies in building a system that can automatically identify sensitive information, namely email, phone number, credit card number, balance, and URL, from the NoSQL query by using an NER model, which falls in the NLP domain; encrypt them by using AES, for efficient search create blind index; generate a vectorized query for securing non-sensitive data; and cluster them for optimization. Among DL algorithms, DBN-LSTM provided the highest performance accuracy (93\%), precision (94\%), recall (93\%), and F1-score (93\%). Whereas database 
Optimization is validated by using the Enron dataset. We find that the proposed system requires less execution time and is more secure than most competing systems.

These approaches, however, have some limitations. Our proposed NER model can identify only a fixed set of entities, as defined by the dataset. Complex queries involving multiple non-sensitive pieces of information are not discussed. Another limitation is space complexity; storing encrypted data, index numbers, and vector values might consume storage. However, a slight increase in storage can be overlooked, as our proposed system offers a double layer of security while preserving optimization through indexing and clustering. Security and efficiency are the utmost requirements, especially when the database has sensitive information.

As part of our future work, we will improve our model by addressing those limitations. Furthermore, implementing complex overlooked aggregations, such as aggregates or joins. Additionally, the proposed scheme currently demonstrates only that storage and retrieval are faster and more secure; in the future, vectorized data retrieval and conversion back to the original form will be implemented. Moreover, we will use more diverse clustering algorithms to improve clustering. The volume of sensitive information obtainable on the web, as well as in businesses and other industries, is increasing. Our proposed model is a strong need to secure sensitive information and personal data, while optimizing.

\bibliographystyle{IEEEtran}
\bibliography{reference.bib}
\end{document}